\documentclass{aastex62}
\usepackage{amsmath}
\usepackage{mathtools}
\usepackage[flushleft]{threeparttable}
\usepackage{lineno}

\graphicspath{{./}{figures/}}

\accepted{July 18, 2022}%{\today}
\submitjournal{ApJ}

\shorttitle{Global modeling of nebulae. III}
\shortauthors{Estrada and Cuzzi}
\begin{document}

\title{Global Modeling of Nebulae With Particle Growth, Drift, and Evaporation Fronts. \\
III. Redistribution of Refractories and Volatiles}

\correspondingauthor{Paul R. Estrada}
\email{Paul.R.Estrada@nasa.gov}

\author{Paul R. Estrada}
\affiliation{NASA Ames Research Center, MS 245-3, Moffett Field, CA 94035}

\author{Jeffrey N. Cuzzi}
\affiliation{NASA Ames Research Center, MS 245-3, Moffett Field, CA 94035}




\defcitealias{Est16}{Paper I}
\defcitealias{Est21}{Paper II}

\begin{abstract}

Formation of the first %100-km size 
planetesimals remains an unsolved problem. %``Simple'' 
Growth by sticking must initiate the process, %and traditionally this incremental growth was seen as continuing to the 100km size range. 
but multiple studies have revealed a series of barriers that can slow or stall %to such incremental 
growth, most of them due to nebula turbulence. %For a while it seemed that the nebula might have been entirely nonturbulent, but this no longer seems to be a viable assumption. Consequently, the highest fidelity models of particle growth in turbulence are needed to see how far it can go as a function of turbulent intensity, location, particle composition, and time. This paper and a companion paper extend this endeavor to global models of particle growth as porous, fractal aggregates.  
In a companion paper, we study the influence of these barriers on models of fractal aggregate and solid, compact particle growth %and transport 
in a %turbulent, 
viscously evolving solar-like nebula %of solar metallicity 
for a range of turbulent intensities $\alpha_{\rm{t}} = 10^{-5}-10^{-2}$. % over 0.5 Myr using our global nebula evolution code. %In that paper, we too focus mainly on the evolution of the particle mass distribution. 
Here, we examine how disk composition %of the disk 
in these same models %, which includes multiple refractory and volatile species, 
changes with time. We find that %the combination of 
advection and diffusion of small grains and vapor, and radial inward drift for larger compact particles and fractal aggregates, 
naturally lead to %implies 
diverse outcomes for planetesimal composition. % as a function of radial location. % in the disk. 
Larger particles can undergo substantial inward radial migration due to %nebula 
gas drag before being collisionally fragmented %destroyed by mutual collisions 
or partially evaporating at various temperatures. This leads to enhancement of the associated volatile in both vapor inside, and solids outside, their respective evaporation fronts, or ``snowlines''. %\textcolor{red}{ not sure what the following means: which can then be transported large distances depending on turbulent intensity.} 
%Conversely, vapor and smaller grains can be diffused or advected outward with the nebula gas.
In cases of 
%For 
lower $\alpha_{\rm{t}}$, we see %that 
narrow belts of volatile or supervolatile material develop in the outer nebula, which could be connected to the bands of ``pebbles" seen by ALMA. Volatile bands, which migrate inwards as the disk cools, can persist %or even build up 
over long timescales as their gas phase continues to advect or diffuse outward across its evaporation front. %, especially in the cases of lowest $\alpha_{\rm{t}}$ where there is more mass in the vapor phase relative to solids. 
%Such locations migrate with time as the disk cools, opening up the possibility of leaving any early-formed planetesimals behind. 
These belts could be sites where supervolatile-rich planetesimals form, such as the rare CO-rich and water-poor comets; giant planets formed just outside the H$_2$O snowline may be enhanced in water. %\textcolor{red}{This is too vague for the abstract, explain or delete: Moreover, they would likely be enhanced in C/O (and potentially C/H) relative to solar once the disk cooled sufficiently.}

\end{abstract}



%\linenumbers

\section{Introduction}\label{sec:intro}

Global models of the evolution of solids in the protoplanetary nebula are critical for understanding the sizes and composition of primitive inner and outer solar system bodies, their accretion timescale, and the meteorite record. Parent bodies of %primitive 
chondrites accreted over a few Myr, a time span over which models and observations of ``dusty-gas'' protoplanetary disks indicate that they %, and the solar luminosity, 
evolved significantly \citep[e.g.,][]{Vil09,Wil12, DM94}. Indeed, growing indications are that the very first planetesimals even form in the first Myr or less \citep{Kru17,Nan19}. 
Understanding how these first %primitive 
bodies formed remains the most intractable question in addressing planet formation, because it is known that even weakly turbulent environments frustrate particle growth with a gauntlet of barriers. These barriers include bouncing and fragmentation \citep{Gut10,Zso10}, which slow or stall growth at pebble sizes, and radially inward migration due to aerodynamic drag \citep{Wei77,CW06,Bra08,Joh14} which can remove particles faster than they can grow. Inward drift can transport material from the outer to inner disk and ultimately into the central star \citep[hereafter  Papers I and II, and other studies noted below]{Est16,Est21}. Moreover, turbulence can even create ``erosion barriers" that prevent incremental growth through the $1-10$ km size range \citep{Idaetal2008,Mor09,NelsonGressel2010,Gresseletal2011}.

%Mitigation of these growth barriers can apparently be achieved when considering very weak or non-turbulent nebulae, or very late-stage disks when most of the nebula gas has dissipated {\it Comment; I'm not sure about this late-stage thing; let's talk about it}. 
These barriers have motivated an emphasis on essentially {\it globally nonturbulent disks} which allow dramatic vertical settling of particles and permit a variety of dense midplane particle layer instabilities  \citep[e.g.,][]{GW73,GP00,GL04}, with the currently most popular being the Streaming Instability \citep[SI,][]{YG05,Joh07, Car17, Yan17} that further concentrates the dense layer of settled particles such that eventual gravitational collapse can ``leap-frog'' clumps of relatively small particles directly  into 100km-size planetesimals. % THIS FEELS LIKE A DISTRACTION TO ME here: it's connected to the idea that the gas can vanish leaving particles of some given size behind. This is not, fr instance, what Birnstiel's models find. The latter may apply most readily to the formation of, e.g., Kuiper Belt single and binary objects \citep[e.g.,][]{Nes19} which though still primordial presumably formed much later after the solar nebula gas had evolved significantly where low gas densities allow even small pebbles to have large Stokes numbers (which is a measure of the influence of the nebula gas on the motions of the particles). 

However, it is increasingly thought that protoplanetary disks in the early stages of their evolution when the first planetesimals and planets form are at least weakly-to-moderately-turbulent in the regions of the disk ($\lesssim 100$ AU) in which particle growth is of the greatest interest \citep[see e.g.,][for reviews]{Tur14,LU19}.  %leap-frog mechanisms such as 
Under such conditions,  formation of planetesimals by SI is  challenged or even precluded under self-consistent particle size assumptions \citep[][\citetalias{Est21}]{Umu20}, at least for disks with cosmic (solar) metallicity. Meanwhile, a different ``leap-frog" mechanism (Turbulent Concentration or TC) does operate in turbulence, and seems capable of producing large planetesimals directly from swarms of growth-frustrated “pebbles” under these conditions \citep{Cuz10, Chambers2010, HC20}. 

The relative efficacy of these alternate mechanisms is strongly dependent on the gas turbulent intensity, the particle size and density, and the {\it local} metallicity, making high-fidelity, self-consistent models of early particle growth and transport in moderate turbulence critical.
%We should squarely face the likelihood that at least the first planetesimals formed in the presence of some level of turbulence (see also below). %, and that their compositional diversity is a reflection of conditions in the protoplanetary disk at the time of their formation.   
Previous global disk models (many of them assuming turbulence at various intensities) have focused on the dynamical evolution of the particle size distribution, and on determining the conditions under which planetesimals can form \citep[][\citetalias{Est16}]{Bra08, Bir10, Dra13}. In these models, relatively little attention has been given to the evolution of the composition of the planetesimals. %\textcolor{blue}{I moved the following here from below:} 
There have also been a number of model studies aimed more specifically at the compositional variations one might expect, some including particle drift and diffusion, and mainly emphasizing the volatiles of the outer solar system \citep{Obe11,AliDib_etal_2014,Piso_etal_2015,Piso_etal_2016,Thiabaud_etal_2015,Obe16,Boothetal2017,Cridland_etal_2017,Bosman_etal_2018,Kri18,Kri20}.  These results resonate with a growing number of observational studies using ALMA and other facilities \citep[and others]{Qi_etal_2013a,Qi_etal_2013b, Cleeves2016, Zhang_etal_2020a, Zhang_etal_2020b, VanDerMarel_etal_2021}. However, the particle growth physics,  radiative transfer, and global mass redistribution in most of these latter models has been significantly simplified to focus on the  chemistry.

The composition of primitive bodies (the mineralogy and isotopic composition of primitive meteorite parent bodies, and increasingly detailed spectroscopy and {\it in-situ} sampling of comets) is rich in constraints on the environments in which they accreted. The rapid inward migration of solids can lead to significant redistribution of the nebula condensibles relative to the evolving gas over time, and alter the local chemical and isotopic composition. Variable C/O ratios affect the local oxidation state, and thus the mineralogy. Remote and in-situ observations of comets suggest that, while a substantial amount of mixing occurred over the solar nebula lifetime \citep{Zolensky_etal_2006, Zanda_etal_2018} - most easily explained by nebula turbulence -  there is also considerable compositional diversity in the comet and TNO populations \citep{Barucci_etal_2008,MummaCharnley2011, Biver_etal_2018,Wierzchos_Womack_2018, McKay_etal_2019}. 
The complex meteorite record  \citepalias[e.g.,][\citealt{Kru17, Nan19, Pignatale_etal_2019, Jacquet_etal_2019, Li20,Li21,Sen21}]{Est16} strongly suggests that even the first few hundred thousand years of nebula evolution (the period covered by this paper) could have left us visible chemical and isotopic fingerprints. %There have been a number of model studies aimed at addressing the compositional variations one might expect, mainly focusing on the influence of particle drift and diffusion, and mainly emphasizing the volatiles of the outer solar system \citep{Obe11,AliDib_etal_2014,Piso_etal_2015,Piso_etal_2016,Thiabaud_etal_2015,Obe16,Boothetal2017,Cridland_etal_2017,Bosman_etal_2018,Kri18,Kri20}.  These results resonate with a growing number of observational studies using ALMA and other facilities \citep[and others]{Qi_etal_2013a,Qi_etal_2013b, Cleeves2016, Zhang_etal_2020a, Zhang_etal_2020b, VanDerMarel_etal_2021}. However, the particle growth physics,  radiative transfer, and global mass redistribution in most of these latter models has been significantly simplified to focus on the  chemistry.  

%The evolving solids may experience local enhancements which may initiate ``leapfrog" planetesimal formation from concentrations of small particles. Processes known to be active in the first fraction of a million years include ongoing infall from the parent cloud and outgoing winds from the sun or disk. Very little attention has been paid to the roles these processes in planetesimal formation, even though evidence now indicates that planetesimals were already forming during this very timeframe.
 
In \citetalias{Est16} we developed our basic techniques and presented global evolutionary models of particle growth (as compact spheres) and drift, with self-consistent radiative transfer and local thermal structure. In our companion \citetalias{Est21}
we present detailed global disk evolution simulations of  porous (fractal) particle aggregate growth by adapting a previously developed particle aggregation and compaction recipe \citep{Suy12,Oku12,Kat13a} to the basic code of \citetalias{Est16}. \citetalias{Est21} focused specifically on the evolution of the particle mass and internal density distribution, whereas here (Paper III) we focus on the radial variation of %{\it bulk} %\textcolor{red}{what do you mean by {\it ``bulk"?}}  
disk composition, and its evolution with time. We do not include a chemical model in our code. Rather, our disk model tracks a realistic mixture of refractory oxides (CAIs), silicates, iron metal and iron sulfide, refractory organics, and ices from H$_2$O to the supervolatiles CO$_2$, CH$_4$, and CO, each with its own associated evaporation front (EF). As in both \citetalias{Est16} and \citetalias{Est21}, we include a self-consistent calculation of disk opacity and temperature which depends on the evolving size distribution in a fully evolving nebula. Our model is novel in its treatment of all this complexity simultaneously. In Section \ref{sec:model} we briefly summarize the most relevant physics for this paper, and refer the reader to   \citetalias{Est21} for more detailed discussion. In Section \ref{sec:results} we describe the results of our simulations in the context of composition, comparing models with fractal aggregation and compaction to compact particle growth models. In Section \ref{sec:discuss} we present a discussion of some implications of our results, and in Section \ref{sec:sum} we summarize our findings.

\section{Nebula model}
\label{sec:model}
 
The simulations presented here and in \citetalias{Est21} are done using our parallelized 
$1+1$D radial nebula code in which we simultaneously treat particle growth and radial migration of solids, 
while evolving the dynamical and thermal evolution of the protoplanetary gas disk. Our code includes the
self-consistent growth and radial drift of particles of all sizes, accounts for size-dependent vertical
settling and diffusion of the particles, radial diffusion and advection of solid and
vapor phases of multiple species of refractories and volatiles, and a self-consistent
calculation of the opacity and disk temperature that allows for us to track the evaporation
and condensation of all species as they are transported throughout the disk. %It should be noted that we only track bulk composition, and that our code does not include a proper chemical model. 
Our base code is described
in detail in \citetalias{Est16}, while the model for fractal growth and its implementation is described in \citetalias{Est21} along with other changes and improvements to our code. Here we only summarize those elements that are germane to the compositional evolution of condensables.
 
\subsection{Gas Disk Evolution}
\label{sec:gasevol}

In our simulations we use initial conditions derived from the analytical expressions of \citet{LP74}, as generalized by \citet{Har98} %for a T Tauri disk \textcolor{red}{maybe change this description, earlier type disk, Desch model-like?} which are 
and parameterized in terms of an initial disk mass $M_{\rm{D}}$ and radial scale factor $R_0$ \citepalias{Est16}. The resulting gas surface density distribution is fairly similar to that used by \citet{Des07} and represents a denser inner nebula than the Minimum Mass Solar Nebula \citep[MMSN,][]{Hay81}. The time-dependent evolution of the gas surface density $\Sigma$ and gas radial velocity $v_{\rm{g}}$ in one dimension are given by \citep{Pri81}

\begin{equation}
    \label{equ:gasevol}
    \frac{\partial \Sigma}{\partial t} = \frac{3}{R}\frac{\partial}{\partial R}\left\{R^{1/2}\frac{\partial}{\partial R}(R^{1/2}\nu\Sigma)\right\};
\end{equation}

\begin{equation}
    \label{equ:vgas}
    v_{\rm{g}} = -\frac{3}{R^{1/2}\Sigma}\frac{\partial}{\partial R}(R^{1/2}\nu\Sigma),
\end{equation}

\noindent
where $\nu = \alpha_{\rm{t}}cH$ is the total viscosity parameterized in terms of a turbulent intensity $\alpha_{\rm{t}}$. The viscosity depends on the evolving disk temperature $T$ both through the gas sound speed $c$, and the nebula gas pressure scale height $H = c/\Omega$ where $\Omega$ is the local orbital frequency. The gas mass accretion rate is given by $\dot{M} = - 2\pi R\Sigma v_{\rm{g}}$ where $\dot{M} > 0$ indicates accretion towards the star, and $\dot{M} < 0$ indicates mass flux outwards. The radius $R_{\rm{t}}$ at which flux reversal occurs depends on ambient nebula conditions and varies with time, tending to move outwards as the disk evolves. Its initial value can be given in terms of the radial scale factor at $t=0$ and is $R_{\rm{t}} \simeq R_0/2$ for our initial conditions \citep{Har98}.  We take $R_0$ as a free parameter which sets the initial surface density profile in terms of the initial disk mass \citepalias[see][]{Est21}.

%\textcolor{red}{mention Desch model?} %The rationale for the chosen value of $R_0$ in the literature has not always been clear, however. A value of $R_0 = 10$ AU was preferred by \citet{Har98} based on young star statistics \citep[a value used by][]{Est16}, but it has also been used to match specifically solar-system-specific angular momentum \citep[$R_0=4.5$ AU,][]{Cuz03}, or set to even larger values, $20-60$ AU \citep{CC06,Gar07,Bra08,HA12,YC12}. 
%We choose $R_0 = 20$ AU and $M_{\rm{disk}} = 0.2$ M$_\odot$ for a central star of mass $M_\star = 1$ M$_\odot$ as our fiducial values, but explore other models with $R_0 = 60$ AU, and different disk masses to show how theses choices influence our model simulations \citepalias[see][]{Est20}.

\subsection{Evolution of Solid and Vapor Phases for Different Compositional Species}
\label{sec:dustvapevol}

Our model tracks multiple species in both vapor and solid phases. In Table \ref{tab:species}, we list the species used in this work along with their corresponding 50\% condensation temperatures $T_i$, compact particle density $\rho_i$ and initial %uniform 
mass fractions in the solid state $\bar{x}_i$, %\equiv \Sigma_i/\Sigma_g$, \textcolor{red}{I added the definition, is it correct? } 
the latter of which are determined with data from Table 2 of \citet{Lod03}. Our nominal species and abundances differ slightly, however, for better alignment with recent studies of comets \citep{MummaCharnley2011}. To get our values for $\bar{x}_i$, we started with the initial refractory organics fraction (CHON, ratio of 1:1:0.5:0.12) from \citet{Pol94} as a guide, and then for simplicity partitioned the remaining C in the ratio of 1:1:1 %\footnote{We make this choice for simplicity, and acknowledge that it may not reflect the breakdown of refractory and gas phase C in the ISM.} 
in determining the fractions of CO, CH$_4$ and CO$_2$. The silicate abundances are initialized by placing all the Mg in orthopyroxene (MgSiO$_3$) and olivine (Mg$_2$SiO$_4$), while the initial refractory iron fraction is what remains after all of the S is placed in FeS. Some of the remaining O is used in determining the Ca-Al fraction which is a mixture of Al$_2$O$_3$, CaO and CaTiO$_3$, while the rest is placed in water. By contrast, \citet{Lod03} has no refractory organics, assumes no O-bearing supervolatiles ({\it i.e.}, CO, CO$_2$), allows water to capture almost all the O that is not consumed by silicates, and assumes the C is in methane ice or in methane hydrate that travels with the water ice. %Despite these differences, our water/rock fraction of 1.15 is similar to the \citet{Lod03} value of 1.17. 
In Table \ref{tab:species}, we also compare the %\sout{absolute} \textcolor{red}{solids-relative?}
mass fractions $F_i \equiv \bar{x}_i/\bar{Z}$ in our work to those used by other workers. Here $\bar{Z} = \sum_i \bar{x}_i$ is the metallicity; for this work $\bar{Z} \simeq 0.014$.

\begin{table}[h!]
\renewcommand{\thetable}{\arabic{table}}
\centering
\caption{List of Species} \label{tab:species}
\begin{threeparttable}
\begin{tabular}{lccccccc}
\tablewidth{0pt}
\hline
\hline

Species & $T_i$ (K) & $\rho_i$ (g cm$^{-3}$) & $\bar{x}_i$ ($T<T_i$) & $F_i$ & $F_i$(L03)\tnote{a} & $F_i$(P94)\tnote{d} & $F_i$(B18)\tnote{f} \\
\hline
Ca-Al & 2000 & 4.0 & $2.38\times 10^{-4}$ & 0.017 & ... & ... & ... \\
Iron & 1810 & 7.8 & $6.54\times 10^{-4}$ & 0.048 & ... & 0.009 & ... \\
Silicates & 1450 & 3.4 & $3.01\times 10^{-3}$ & 0.220 & (0.204)\tnote{b} & 0.244 & 0.329\tnote{g} \\
FeS & 680 & 4.8 & $1.16\times 10^{-3}$ & 0.085 & (0.125)\tnote{c} & 0.055 & 0.074 \\
Organics & 425 & 1.5 & $3.53\times 10^{-3}$ & 0.258 & ... & 0.252\tnote{e} & 0.397 \\
H$_2$O & 160 & 0.9 & $3.46\times 10^{-3}$ & 0.253 & 0.384 & 0.397 & 0.200 \\
CO$_2$ & 47 & 1.56 & $8.16\times 10^{-4}$ & 0.060 & ... & ... & ... \\
CH$_4$ & 31 & 0.43 & $2.99\times 10^{-4}$ & 0.022 & 0.222 & ... & ... \\
CO & 20 & 1.6 & $5.19\times 10^{-4}$ & 0.038 & ... & ... & ... \\
 
\hline
%\multicolumn{5}{c}{NOTE. - Two decimal aligned columns}
\end{tabular}
 \begin{tablenotes}
  \item[a] Table 11, first column of \citet{Lod03}
  \item[b] This value includes silicates and oxides. %``Rock'' includes this, as well as all metals + FeS  \textcolor{red}{The legend is a little confusing. You only use the term ``rock" ONCE, just above this table. Do you mean that for Lodders, ``rock" = silicates + oxides + metals + FeS? Maybe when you give the ice/rock number above, give a footnote saying what we, and what Lodders, assume ``rock" is specifically. And "oxides" is confusing here because the Ca-Al "CAIs" are oxides. By "metals" in (b) do you mean Iron + FeS? or maybe all siderophiles? Which is probably what our "Iron" is. Then, below in (c) you say Lodders FeS includes "all metals", which (eg.) includes Ca-Al materials? Aren't these the oxides mentioned under silicates in (b)? Just need a little more clarity, I think.}
  \item[c] Includes all metals (e.g., refractory Fe) in addition to FeS
  \item[d] Table 2, third column of \citet{Pol94}
  \item[e] Does not include ``volatile organics'' such as CH$_3$OH and H$_2$CO used in \citeauthor{Pol94}
  \item[f] Table 1, third column of \citet{Bir18}
  \item[g] Astronomical (amorphous) silicates from \citet{Dra03}
 \end{tablenotes}
\end{threeparttable}

%\caption{\textcolor{red}{Not trying to drive you crazy but I think for our own future reference and Diane's use, AND for the reader who might appreciate the differences, maybe you can add a few more columns of $x_i$, for Lodders, the DSHARP team (Birnstiel et al 2018 I think), and Pollack. Most of the difference is that these guys will have zeroes in some lines and different values in others. Birnstiel's nominal is only 1/3 of Lodders' water ice and NO SVs.}}
\end{table}

As the disk evolves, the solids  ({\it  dust} and larger, {\it migrator} particles,  see below) and {\it vapor} fractions, denoted by d, m and v respectively, of each species change with time as particles grow and are transported throughout the nebula. The local instantaneous mass fraction, or concentrations $x_i$ of each species $i$ relative to the gas, in each phase, are given as  functions of time by
\begin{equation}
    \label{equ:alphas}
    x_i^{\rm{v,d,m}} \equiv \frac{\Sigma_i^{\rm{v,d,m}}}{\Sigma},
\end{equation}

\noindent
or the ratio of the surface density of each species and phase  to the gas surface density. The mass fraction of all species in a given phase  $f_{\rm{v,d,m}}$ is just the sum over $i$ in Eq. (\ref{equ:alphas}), e.g. for the dust, $f_{\rm{d}} = \sum_i x^{\rm{d}}_i$. Similarly, the local varying mass fraction for all phases of a given species is the sum over all {\it i}, $x_i^{\rm{v}} +x_i^{\rm{d}} +x_i^{\rm{m}}$, whereas the instantaneous metallicity, which includes all species in the solid state locally, %in our simulations 
is $Z = \sum_i ( x_i^{\rm{d}} + x_i^{\rm{m}})$.

Particle growth in our code incorporates the physics of fragmentation, bouncing, erosion, and mass transfer \citepalias[see][and references therein]{Est16,Est21} with mass, composition,  and impact velocity-dependent sticking coefficients. Our code employs the moments method for the ``dust'' distribution \citep{EC08,EC09} which assumes a powerlaw dependence ranging in mass from a monomer  up to the fragmentation mass, and treats particle growth explicitly beyond that. This allows us to greatly speed up calculations and focus on the more interesting evolution of larger aggregates. A migrator then is a particle that is more massive than the fragmentation mass, tends to be fairly decoupled from the gas phase \citepalias{Est16}, and their distribution is tracked individually by mass rather than as only part of a powerlaw. Examples of the particle mass distributions of many of the simulations used in this paper can be found in the Appendix of \citetalias{Est21}.  The minimum size in our particle size distributions is a monomer of radius $0.1 \mu$m, which has a mass of $6.4\times 10^{-15}$ g when using the compact particle density of $1.52$ g cm$^{-3}$ derived from the combination of all species listed in Table \ref{tab:species}.

In our simulations we use a threshold fragmentation specific energy $Q_{\rm{f}}=10^4$ cm$^2$ s$^{-2}$ for silicates and more volatile ices than water (CO$_2$, CH$_4$, and CO) which we will refer to as {\it supervolatiles} \citep{Mus16}, and adopt $10^6$ cm$^2$ s$^{-2}$ for water ice, which has been thought to be ``stickier'' than silicates \citep[e.g.,][]{Wad09,Oku12}. By stickier, it is meant that much higher impact speeds are needed for fragmentation. Recent work suggests it may only be sticky over a limited temperature range near the condensation temperature of water ice \citep{MW19}, thus we include simulations in which we account for this by adopting the silicate value of $Q_{\rm{f}}$ outside this range which we refer to as a ``cold H$_2$O ice'' model \citep[see Paper II and][for more discussion]{Umu20}. %\citepalias[see][for more discussion]{ Est21}, as well as \citet{Umu20}. 

The evolution of solid and vapor phases is determined for each species by the advection-diffusion equation \citep{Des17}

\begin{equation}
    \label{equ:diffadv}
    \frac{\partial \Sigma_i}{\partial t} = \frac{1}{R}\frac{\partial}{\partial R}\left\{R{\mathcal{D}}\Sigma
    \frac{\partial \alpha_i}{\partial R} - R\bar{v}\Sigma_i\right\} + {\mathcal{S}}_i,
\end{equation}

\noindent
where $\bar{v}$ is the net, inertial space velocity (advection + drift), %\textcolor{red}{I guess ``inertial space advection" must be the net of getting carried with the gas, and headwind drift relative to the gas. Might be worth a comment. Indeed, this would be a good place to define headwind drift and $\beta$, which I do not think IS defined before we hit it in figure 1. It is defined much later on, at the end of 3.1 now. But what really confused me was that below,  it seems $\mathcal{S}_i$ includes radial drift as a source or sink? I would think drift is part of the grad $\bar{v}$ term. }
${\mathcal{D}}$ is the diffusivity and ${\mathcal{S}}_i$ represents sources and sinks for species $i$ which include growth, radial transport and destruction of migrating material. For the vapor phase, $\bar{v} = v_{\rm{g}}$ and ${\mathcal{D}}=\nu$, whereas for the particles $\bar{v}$ and ${\mathcal{D}}$ depend  on a particle's stopping time\footnote{The stopping time $t_{\rm{s}}$ is the time needed for gas drag to dissipate a particle's momentum of mass $m$ and radius $r$ relative to a gas with local volume density $\rho$.} $t_{\rm{s}}$, usually expressed as the Stokes number ${\rm{St}} =  t_{\rm{s}}\Omega$. $\bar{v}$ for the particles has two contributions, one imposed by the radial motion of the gas $v_{\rm{g}}$, and the second the radial velocity of the particle with respect to the gas, which depends on the normalized pressure gradient or headwind parameter 

\begin{equation}
    \label{equ:pgrad}
    \beta(R,t) = -\frac{1}{2\rho\Omega v_{\rm{K}}}\frac{\partial p}{\partial R} = -\frac{1}{2}\left(\frac{c}
    {v_{\rm{K}}}\right)^2\frac{\partial \,{\rm{ln}}\,p}{\partial \,{\rm{ln}}\, R},
\end{equation}

\noindent
%$\beta \propto (c/v_{\rm{K}})^2$ 
where $p$ is the pressure, $\rho$ is the nebula gas mass volume density, $c$ is the gas sound speed, and $v_{\rm{K}}$ the local Kepler velocity (see Sec. \ref{sec:evoldisk}). Depending on size and density, and local gas conditions, particles can be subject to different drag regimes that determine $t_{\rm s}$ \citepalias[see ][ Sec. 2.2.1]{Est21}. In the end, $\bar{v}$ and ${\mathcal{D}}$ are determined at any radius $R$ through a mass density-weighted mean over all particle masses. %\textcolor{red}{I rewrote the above few  sentences}
In Equation \ref{equ:diffadv} above, the sign of the radial drift velocity for solids depends on particle mass. Massive particles tend to drift radially inwards ($\bar{v} < 0$), while less massive ones can be radially advected outwards with the gas ($\bar{v} > 0$). This effect is accounted for in our code for particles smaller than the fragmentation mass by determining the particle mass (where $\bar{v} \approx 0$) that separates the population of particles moving inward from that moving outward, and then using these  relative mass fractions as weighting factors in solving the advection-diffusion equation for each population. For masses larger than the fragmentation mass, we account for drift (and diffusion) for all mass bins individually \citepalias[see][]{Est16,Est21}.

\subsection{Disk Thermal Evolution}
\label{sec:disktemp}

The evolving size distribution affects the local disk temperature through the local opacity, so a self-consistent calculation of the disk temperature is essential to capturing the disk's dynamical evolution. The midplane temperature $T$ is determined iteratively from

\begin{equation}
    \label{equ:temp}
    \sigma_{\rm{SB}}T^4 = \frac{9}{8}\nu\Sigma\Omega^2\left(\frac{3\tau_{\rm{R}}}{8}+\frac{1}{2\tau_{\rm{P}}}
    \right) + \frac{L_\star \phi}{4\pi R^2},
\end{equation}

\noindent
which represents a combination of internal heating due to viscous dissipation  and external illumination by the stellar luminosity $L_\star$. The optical depths $\tau_R$ and $\tau_P$ are the Rosseland and Planck means, applied in optically thick and thin regions respectively. We assume a flared disk geometry with a general radial variation given by \citet{CG97} \citep[see, also][] {KH87,RP91}:
\begin{equation}
    \label{equ:graze}
    \phi(R) \sim 0.005 R^{-1}_{\rm{AU}} + 0.05 R^{2/7}_{\rm{AU}}.
\end{equation}

\noindent
Our simulations employ a time variable stellar luminosity using a model for a 1 M$_\odot$ star by  \citet[][also \citealt{Sie00}]{DM94}, %\textcolor{red}{were there more recent refs or is this it?} 
where the initial luminosity is roughly $L \approx 12 \,{\rm{L}}_\odot$ at the beginning of a simulation, and drops to $\approx 3 \,{\rm{L}}_\odot$ after 0.5 Myr. Such a high luminosity means that EFs for the supervolatile species listed in Table \ref{tab:species} are initially located considerably further away from the star than would be suggested by equilibrium temperatures calculated using a main sequence luminosity. In our models then, the water snowline begins at around $\sim 10-15$ AU, similar to the post buildup stage in models that include infall \citep[e.g.,][]{DD18,HN18}. We will consider other evolutionary track models in the future \citep[e.g., see][]{DiC09,Tog11}.

Using a temporally constant $\phi$ as given in Eq. \ref{equ:graze} while employing a variable luminosity is a simplification, because the varying luminosity may be primarily due to a variable stellar size. The protostellar flux on the disk surface is the product $L_\star \phi$ (Eq. \ref{equ:temp}) and the effective value of $\phi$ may increase along with $L_\star$  going backwards in time due to the increase in stellar radius. The net effect would probably be an amplification of the temporal decrease of $L_\star \phi$ over time as the luminosity decreases. Since the assumed value for $\phi$ from \citet{CG97} assumes a stellar radius at the small end of our range, this might imply a slightly warmer disk at the earliest times than we currently find, leading to some temporal shift in the radial locations of the various EFs. However, since the values of $L_\star$ and $\phi$ are themselves not that well known, and the disk temperature only depends on the product of them to the 1/4 power, even where it would provide the dominant heating (in the outer nebula), this variation is unlikely to be a major effect.

%which must be solved iteratively because the Rosseland and Planck mean optical depths $\tau_{\rm{R}}$ and $\tau_{\rm{P}}$, respectively, are temperature dependent and thus affect the evolving particle size distribution and the solids and vapor fractions of all species. In Eq. (\ref{equ:temp}), it is assumed that the first term on the RHS is a local, vertically integrated viscous dissipation rate, while the second term on the RHS accounts for the stellar flux on each disk face, and $\phi$ is a grazing incidence angle depending on disk geometry. In our model, we assume a flared disk geometry with a general radial variation given by 

The optical depths $\tau_R,\tau_P$ in Eq. (\ref{equ:temp}) are functions of the temperature-dependent opacity $\kappa$ as $\tau \equiv \kappa\Sigma/2$. We define the Rosseland and Planck mean opacities from the basic wavelength-dependent opacity $\kappa_\lambda$  in the standard way as:

\begin{equation}
    \label{equ:kappaR}
    \kappa^{-1}_{\rm{R}} = \frac{\pi}{4\sigma_{\rm{SB}}T^3}\int \kappa^{-1}_\lambda\frac{dB_\lambda}{dT}\,d\lambda;
    \,\,\,\,\kappa_{\rm{P}} = \frac{\pi}{\sigma_{\rm{SB}}T^4}\int \kappa_\lambda B_\lambda d\lambda.
\end{equation}

\noindent
To determine the $\kappa_\lambda$ for both compact and porous aggregate particles, we utilize the opacity model of \citet{Cuz14}  which contains realistic material refractive indices for the species listed in Table \ref{tab:species} from \citet{Pol94} and is easily modified to handle porous aggregates, while for the supervolatiles we adopt available values from the literature \citep{War86,Sch89,Hud93,MO94}. For the dominant particle sizes and mid-IR wavelengths, scattering can be neglected. 

\subsection{Mass Transport Across Evaporation Fronts}
\label{sec:EFs}

Evaporation fronts (EFs) are locations in the disk where species-dependent phase changes between solids and vapor can occur, and these play an important role in the redistribution of refractories and volatiles that is the focus of this paper. Inwardly migrating particles can sublimate\footnote{It is possible for, e.g., iron or silicates \citep[e.g.,][]{Nag94} to be stable as a liquid under nebula pressures somewhat higher than ours (our nebula pressures inside 1 AU are in the $\sim 10^{-4}-10^{-3}$ bar range), but we do not include this complication in this work. How this would affect the stickiness of these species may be an important wrinkle to explore in the future.} at various EFs, enhancing the gas phase there, while subsequent outward diffusion of vapor back across the EF can recondense onto grains there -  significantly enhancing both the amount and composition of the resident material \citepalias[e.g.,][\citealt{RJ13,SO17}]{Est16}. These processes have implications for chemistry, minerology, and planetesimal formation. We allow for the ``phase'' change to occur linearly over a small temperature range 
%$T_i - \Delta T_{\rm{EF}} \le T \le T_i + \Delta T_{\rm{EF}}$, with $\Delta T_{\rm{EF}} = 0.05$ K 
($\sim 1$ K) so that EFs are not necessarily sharp boundaries. Allowing for this gradual radial transition prevents unrealistic drops in opacity and temperature just inside an EF, and effectively mimics buffering of midplane  temperature changes as material is evaporated or condensed at different {\it altitudes}  \citepalias[see][for more discussion]{Est16}. %We plan to address the details of phase changes in future work. %\textcolor{red}{add references to Schoonover and Ormel; Ros or Lambrechts and Johansen?}.

The specifics of how changes in the mass fractions $x^{\rm{d,m,v}}_i$ are treated as they are transported across EFs either through radial drift, advection or diffusion are described in Appendix A.3 (and Secs. 2.4.5-2.4.6) of \citetalias{Est16}. In our code, when inwardly transported solid material encounters an EF, %\sout{all (or some fraction if the EF extends over a radial distance larger than a radial bin) of} 
their associated volatile is removed from the solid phase and transferred to the gas phase. 
%\sout{instantaneously}. 
As a by-product of the change in $x_i$, the composition and masses of the particles and aggregates change. When material drifts inwardly  into a new radial bin the released, volatile-free dust fraction is assumed to adjust to the local dust distribution, while the distribution of newly volatile-free migrators is explicitly added to the local population.

When vapor of the associated volatile diffuses outward across its EF, we assume that it recondenses instantaneously, and is partitioned onto both dust and migrator populations in the appropriate area fractions. Changes in the mass distribution affect the local opacity and temperature so that additional modifications to the local $x_i$ can occur when the new $T$ is calculated. 
%\sout{We note that our model is currently not sophisticated enough to account for {\it where} the condensate is deposited on an aggregate (e.g., over its surface),} 
Since most of the outwardly diffused  volatile material condenses on the tiny grains with the most surface area, which can then coagulate with existing aggregates, the recondensed volatile  is assumed to be uniformly mixed back into the local  particles\footnote{ Our particles are treated as if they are aggregates of chemically distinct monomers.}. We plan to address the details of phase changes, as well as more realistic models for particle architecture (see Sec. \ref{sec:cavfwk}) in future work.

%\noindent Note that in the two column style figures and tables will only
%span one column unless specifically ordered across both with the ``*'' flag,
%e.g. \\
%
%\noindent{\tt\string\begin\{figure*\}} ... {\tt\string\end\{figure*\}}, \\
%\noindent{\tt\string\begin\{table*\}} ... {\tt\string\end\{table*\}}, and \\
%\noindent{\tt\string\begin\{deluxetable*\}} ... {\tt\string\end\{deluxetable*\}}. \\
%
%\noindent This option is ignored in the onecolumn style.
 
%Two style options that are needed to fully use the new revision tracking
%feature, see Section \ref{sec:highlight}, are {\tt\string linenumbers} which 
%uses the lineno style file to number each article line in the left margin and 
%{\tt\string trackchanges} which controls the revision and commenting highlight
%output.

%There is also a new {\tt\string modern} option that uses a Daniel
%Foreman-Mackey and David Hogg design to produce stylish, single column
%output that has wider left and right margins. It is designed to have fewer
%words per line to improve reader retention. It also looks better on devices
%with smaller displays such as smart phones.
 
\begin{table}[h!]
\renewcommand{\thetable}{\arabic{table}}
\centering
\caption{List of Simulations} \label{tab:models}
\begin{threeparttable}

\begin{tabular}{lcccc}
\tablewidth{0pt}
\hline
\hline

\tnote{a} Model & Type & $\alpha_{\rm{t}}$ & $Q^{\rm{ice}}_{\rm{f}}$ (cm$^2$ s$^{-2}$) \\
\hline
fa2g & fractal & $10^{-2}$ & $10^6$ \\
\tnote{b} fa3g & fractal & $10^{-3}$ & $10^6$  \\
fa3Qg & fractal & $10^{-3}$ & $10^4$  \\
fa4g & fractal & $10^{-4}$ & $10^6$  \\
fa5g & fractal & $10^{-5}$ & $10^6$  \\
sa2g & compact & $10^{-2}$ & $10^6$  \\
\tnote{b} sa3g & compact & $10^{-3}$ & $10^6$  \\
sa3Qg & compact & $10^{-3}$ & $10^4$  \\
sa4g & compact & $10^{-4}$ & $10^6$  \\
sa5g & compact & $10^{-5}$ & $10^6$  \\

\hline
%\multicolumn{5}{c}{NOTE. - Two decimal aligned columns}
\end{tabular}
\begin{tablenotes}
 \item[a] All models have an initial disk mass of $M_{\rm{disk}} = 0.2$ M$_\odot$, and $R_0 = 20$ AU.
 \item[b] These are the fiducial models from \citetalias{Est21}.
\end{tablenotes}
\end{threeparttable}
\end{table}

\section{Results} 
\label{sec:results}
%\textcolor{red}{Looks like you already added a plot of St as a function of SMA, nice going, probably have made other changes too. In many other cases, where we refer back to Paper I or II, we should cite a specific figure or figures.}
We analyze a specific set of compact and fractal aggregate growth models for different turbulent intensity $\alpha_{\rm t}$ (Table \ref{tab:models}), originally presented in \citetalias{Est21}, where all simulations have $R_0 = 20$ AU and $M_{\rm{disk}} = 0.2$ M$_\odot$ for a central star of mass $M_\star = 1$ M$_\odot$ with variable luminosity (Sec. \ref{sec:disktemp}). In our analysis here, we focus on the compositional evolution. % of the solids. 
We emphasize again that our models do not incorporate a proper chemical model, and thus we only follow the composition and phase of our initial set of condensable species over time. When we refer to the `inner' and `outer' disk as we often do in this paper, we mean {\it inside} and {\it outside} the water snowline, respectively. Details of the evolution of the particle mass, size, and porosity distributions  can be found in \citetalias{Est21}.  
 
\begin{figure}
\gridline{\fig{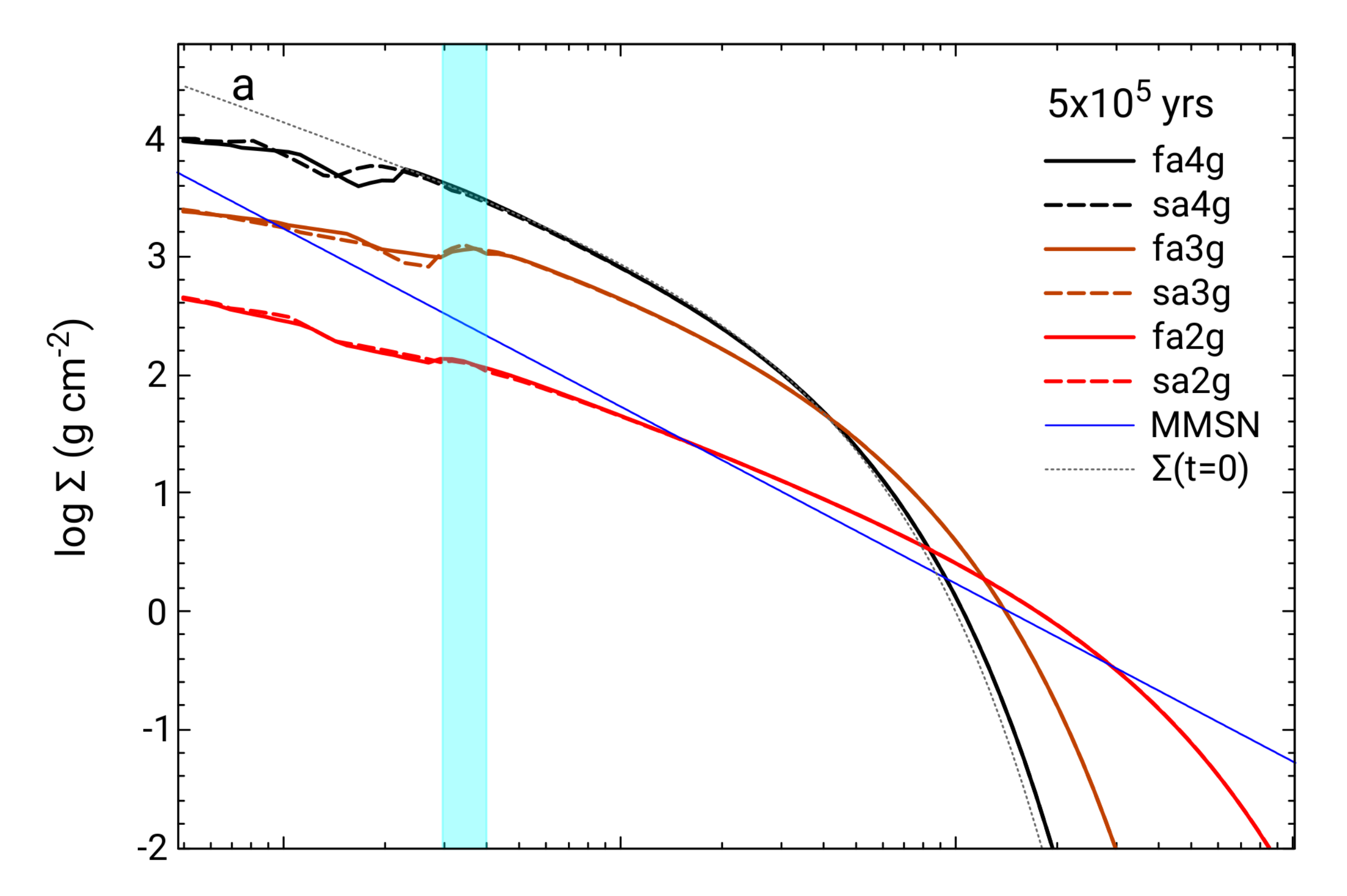}{0.45\textwidth}{}
          \fig{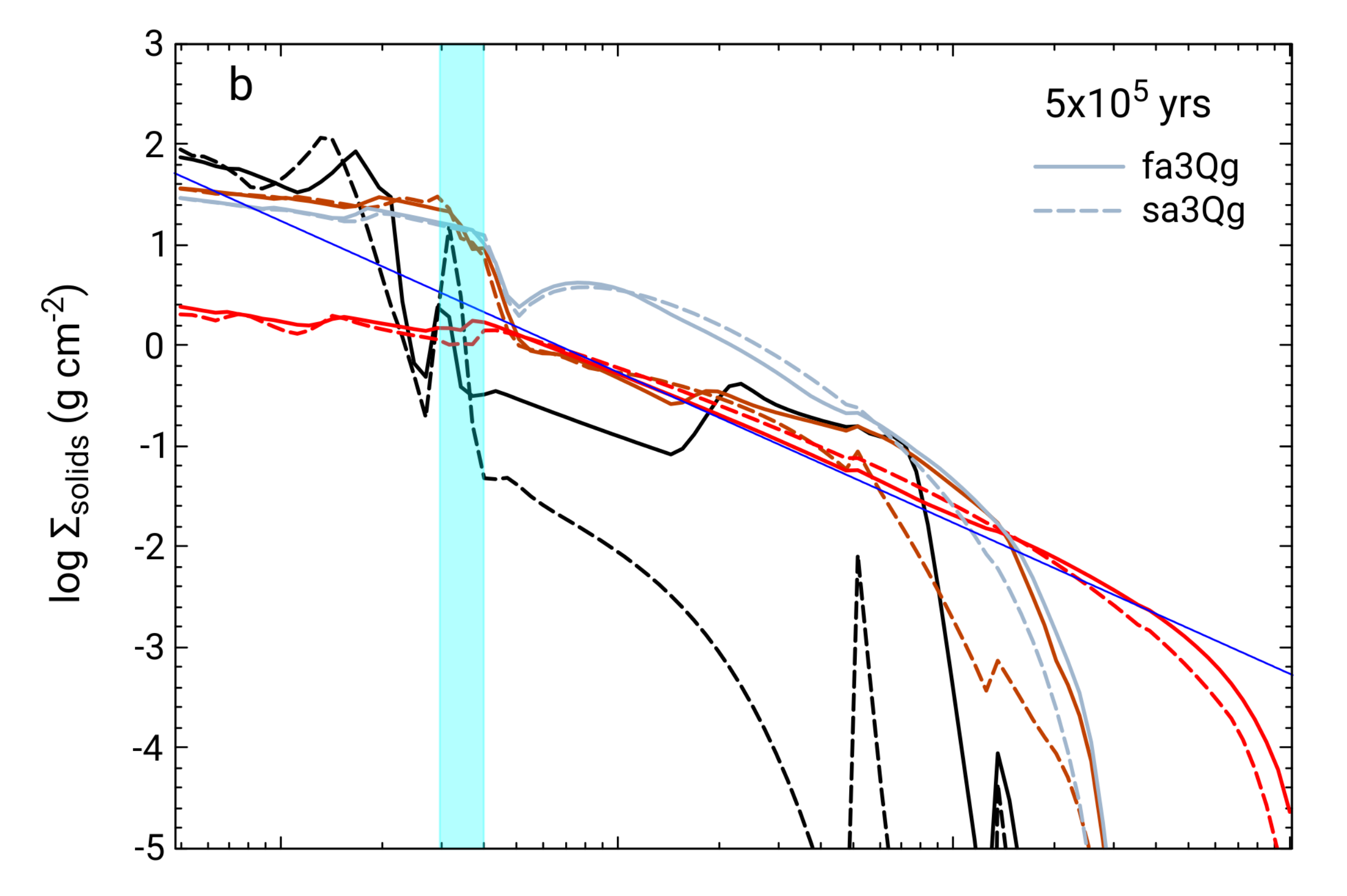}{0.45\textwidth}{}
          }
\vspace{-0.4in}
\gridline{\fig{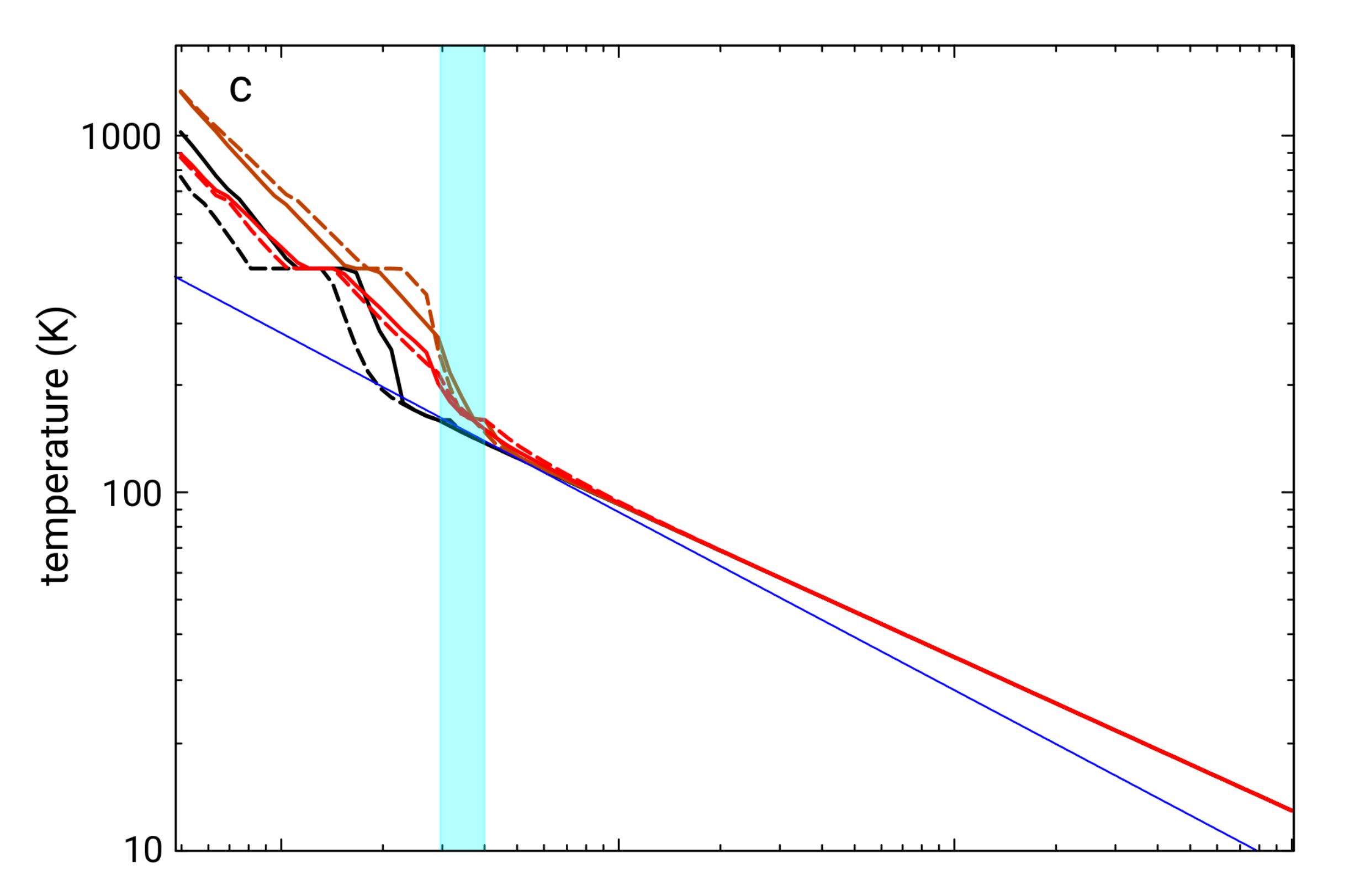}{0.45\textwidth}{}
          \fig{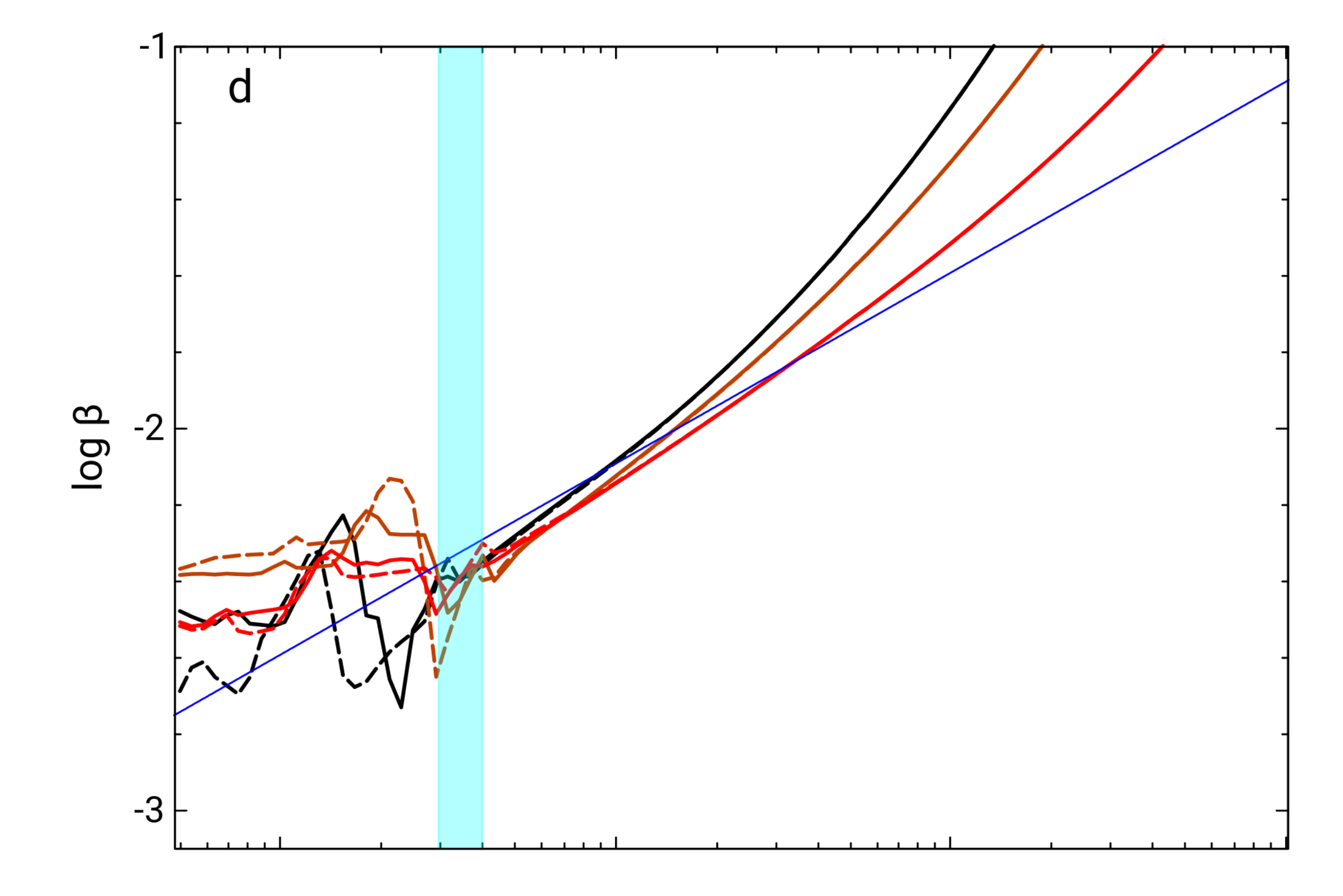}{0.45\textwidth}{}
          }
\vspace{-0.4in}
\gridline{\fig{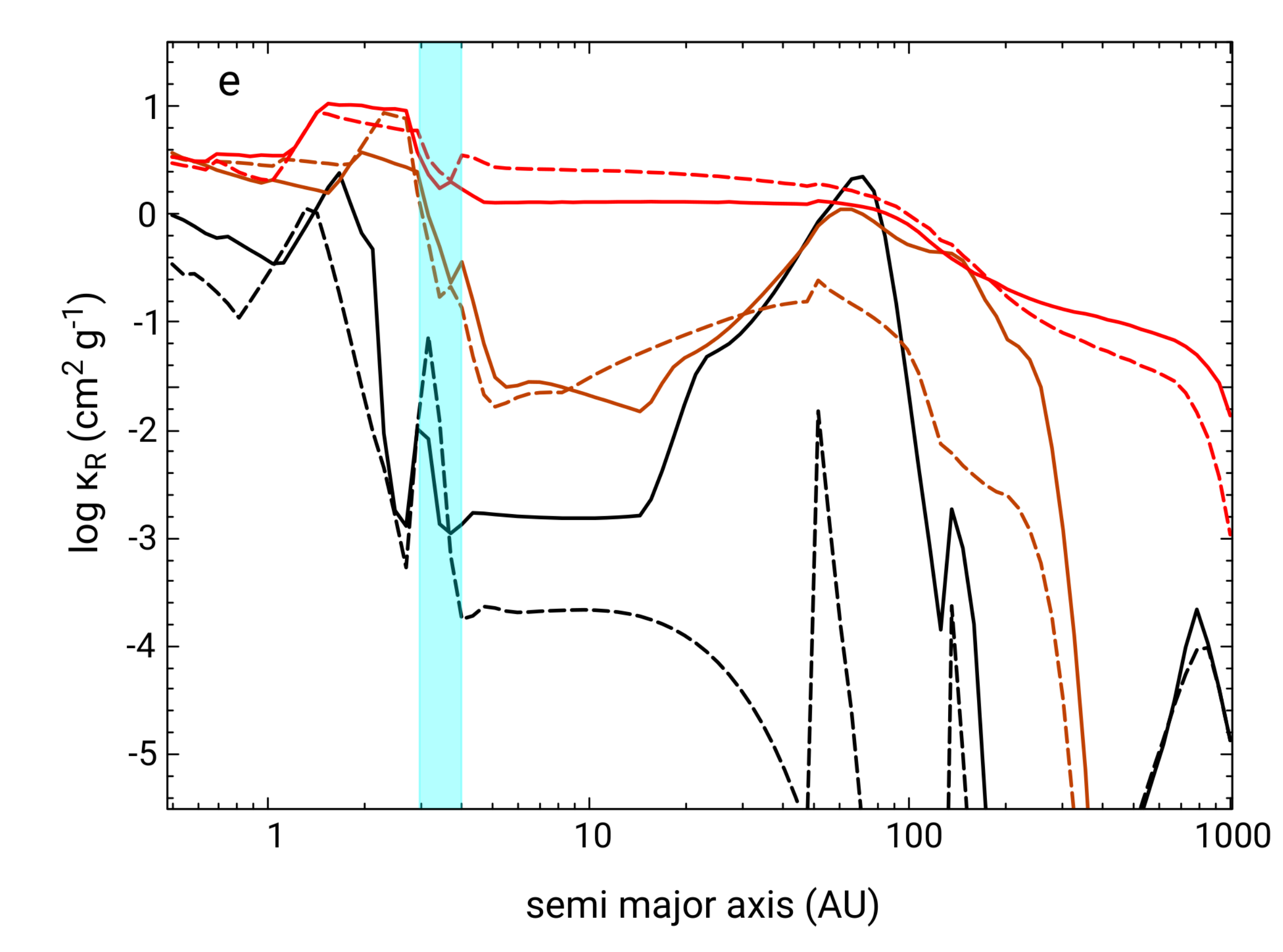}{0.45\textwidth}{}
          \fig{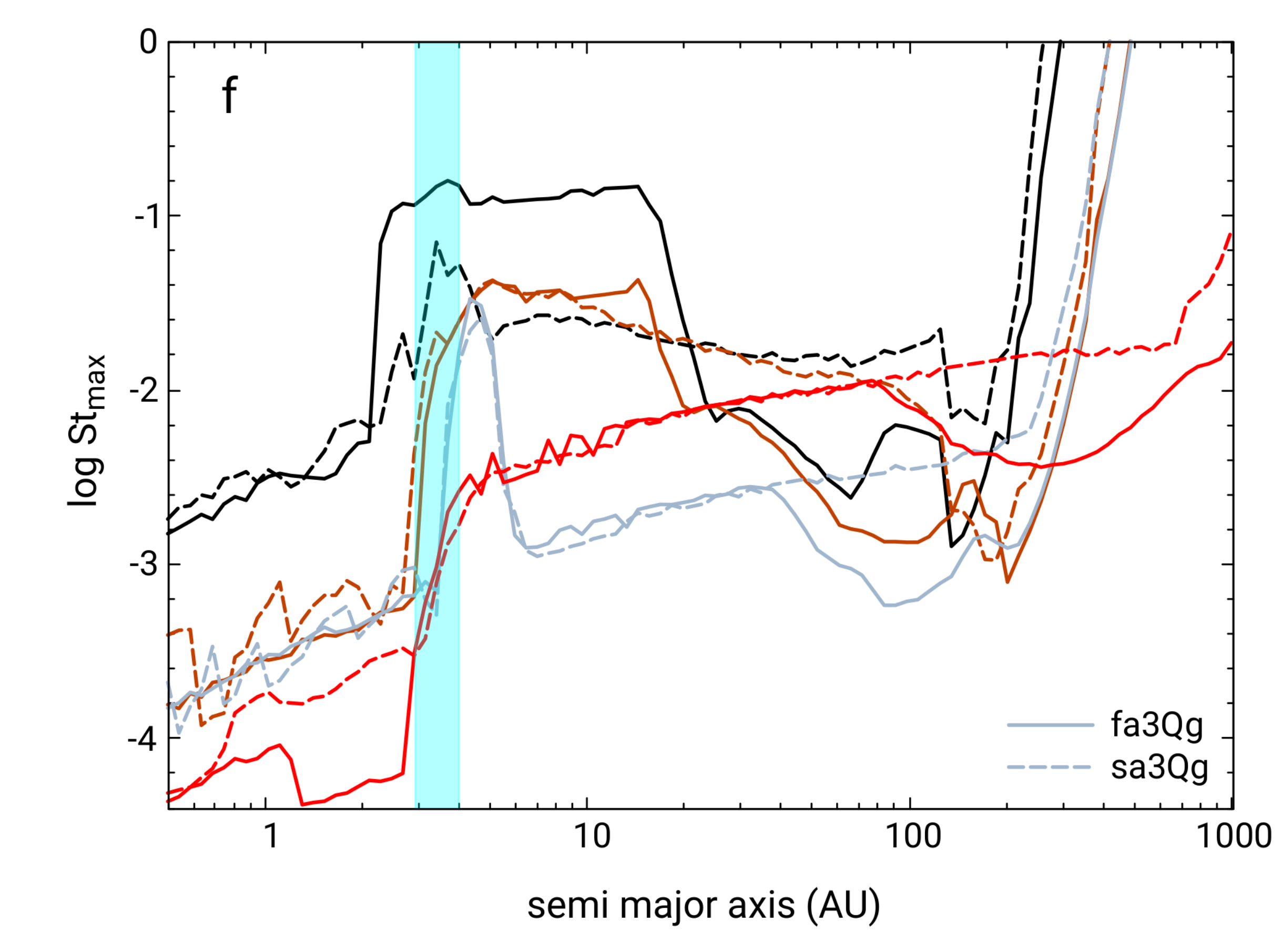}{0.45\textwidth}{}
          }
\caption{Array of ambient disk properties as a function of semi-major axis after 0.5 Myr for different values of the turbulent parameter $\alpha_{\rm{t}}$ as labeled by their model designation (Table \ref{tab:models}). The cyan region shows the radial range of the water ice snowline at this epoch across the models. (a) Gas surface densities. Also plotted is the initial profile at $t = 0$ (grey dotted curve). The blue curve (here and panels b-d) corresponds to the MMSN profile. (b) Solids surface densities for the models shown in panel (a), plus two additional models for cold H$_2$O ice (solid and dashed grey curves). (c) Midplane temperature for the models shown in panel (a). The standard MMSN temperature profile is steeper than our outer disk temperature profiles which are more akin to a passive disk model \citep{CG97}. %\textcolor{red}{isn't the MMSN passive at these distances? Maybe its not flared.} 
(d) Headwind parameter  $\beta$ %\textcolor{red}{have not yet defined, see comment previously} 
for the models shown in panel (a). (e) Rosseland mean opacity for models shown in panel (a). (f) Stokes number ${\rm{St}_{max}}$ of the mass dominant particle or aggregate,  for models shown in panels (a) and (b).  %Panels b and f also plot the $\Sigma_{\rm{solids}}$ and ${\rm{St}_{max}}$ for the ``cold H$_2$O ice'' models (grey curves). 
 %\textcolor{red}{the Rosseland and Planck opacities are SO similar they could almost be combined to emphasize that, maybe show fewer time steps. Then use a panel for the Stokes numbers? or for radial velocities? Or masses? Or pressures? You refer to all these things in the paper.} - ONLY SHOWING ROSSELAND. ST SHOWN. NOT SURE VELOCITIES ARE THAT INFORMATIVE
These models showcase the variation seen across different turbulent intensities: in particular, as $\alpha_{\rm{t}}$ decreases, radial drift more rapidly removes material from the outer disk to the inner disk because particles or aggregates can grow larger (higher St). However, in the fractal models the bulk of material loss is restricted to the region from the water snowline out to $\sim 20$ AU, because outside this region aggregates remain fluffy enough (smaller St than their compact particle counterparts, panel f) that their radial drift remains lower.
\label{fig:diskevol}}
\end{figure}  

\subsection{Evolution of Ambient Disk Properties}
\label{sec:evoldisk}

In Figure \ref{fig:diskevol} (panels a-f) we show a snapshot for both fractal aggregate (solid curves) and compact particle (dashed curves) growth models of the most relevant subset of disk properties after 0.5 Myr of evolution for turbulent intensity values of $\alpha_{\rm{t}}=10^{-4}$ (black curves), $10^{-3}$ (orange curves, which are our fiducial models from \citetalias{Est21}) and $10^{-2}$ (red curves). The cyan region in each plot denotes the radial range over which the water snowline is located across all models ($\sim 2.9-4$ AU) at this evolution time. The top left panel (a) shows the gas surface density $\Sigma$. All models have the same initial $\Sigma$ profile defined by $M_{\rm{disk}}$ (indicated by the light grey dotted curve) %\textcolor{red}{similar comment as in paper II; light grey curve is nearly invisible in hard copy and/or marginal light. Maybe black dotted? Also, as in paper II, I'd prefer solid curves for solid particles and dashed curves for fractal particles, but this might be a lot of work.} 
and are characterized by a steep dropoff outside of $R_0$ \citep{Har98}. These can be compared to the standard MMSN surface density profile ($\Sigma \propto R^{-3/2}$; the blue line) with a disk gas mass an order of magnitude smaller than our $M_{\rm{disk}}$  \citep{Hay81}. %Our initial profiles contain much more mass than the MMSN in the planet forming region between $\sim 2-20$ AU, and less inside and outside these regions. 
Naturally, the low $\alpha_{\rm{t}}$ case shows the least evolution except in those regions where the variation in $\Sigma$ is driven by strong variations in $T$ (panel c) through the viscosity, and radial variations in $\Sigma$ tend to be smoothed out with higher values of $\alpha_{\rm{t}}$. By 0.5 Myr, the gas surface density for $\alpha_{\rm{t}}=10^{-2}$ is well below the standard MMSN value due  to being lost inwards to the star, and to rapid spreading outwards beyond the radius where the mass flux changes sign (at $t=0$, this is $\sim R_0/2$, but this location tends to move outwards with time). 

More interesting variation is seen in the solids surface density $\Sigma_{\rm{solids}}$, and between fractal aggregate and compact particle growth models (panel b). Generally, particle growth leads to inward radial drift, as particles or aggregates decouple more from the gas phase and incur headwind-drag forces. Inward drifting particles eventually  encounter an EF which can lead to vapor enhancement inside, and (if the sublimated species diffuses or is advected back across the EF) solids enhancement outside of it. This effect (and contrast) is seen most easily for $\alpha_{\rm{t}}=10^{-4}$ at the H$_2$O snowline at $\sim 3$ AU. It is also seen for $\alpha_{\rm{t}}=10^{-3}$ but diffusion is stronger so the contrast is less outside the water snowline, but also inside due to the outward diffusion from the organics EF ($\sim 1-2$ AU, more below). With higher $\alpha_{\rm{t}}=10^{-2}$, contrasts are mostly blurred out across all EFs. 

Interior to the snowline (the inner disk), the differences in $\Sigma_{\rm{solids}}$ between fractal and compact growth models for a given $\alpha_{\rm{t}}$  are surprisingly not very  significant, in spite of the fact that the fractal aggregate masses are orders of magnitude larger than their compact counterparts \citepalias[see][]{Est21}. This suggests that their drift rates are similar despite the disparities in mass, and indeed the Stokes numbers (panel f, see below) are similar for $\alpha_{\rm{t}} = 10^{-4}$ and $10^{-3}$. The most variation is seen in the $\alpha_{\rm{t}}=10^{-4}$ case which is due to the aforementioned organics EF which is much broader, and more pronounced than the snowline. This is actually still the case for $\alpha_{\rm{t}}=10^{-3}$; it is less noticeable here in $\Sigma_{\rm{solids}}$ but can be seen clearly in $Z$ (see next section). Stokes numbers for the highest turbulent intensity case are  %\sout{somewhat different from each other, but quite small}, \textcolor{red}{
generally even smaller than for the other cases so these particles remain much more coupled to the nebula gas. 

Outwards of the snowline %\sout{and beyond} 
where the fragmentation energy threshold $Q_{\rm{f}}$ is much larger, things begin to diverge because both fractal and compact growth can proceed to much larger masses (and in fact many orders of magnitude larger for fractal aggregates, see \citetalias{Est21}). This means these particles and aggregates are subject to much more rapid radial drift, owing to their larger Stokes numbers (panel f), leading to marked drops in $\Sigma_{\rm{solids}}$, which is even more evident in the local metallicity $Z$ (see Sec. \ref{sec:evosolvap}). The drop in $\Sigma_{\rm{solids}}$ characterizes both compact and fractal growth models for $\alpha_{\rm{t}}=10^{-4}-10^{-3}$; however, the drop in $\Sigma_{\rm{solids}}$ extends only to $\sim 20$ AU in the fractal models, whereas the compact growth models continue to systematically lose material from the entire outer disk. Even though fractal aggregates grow more massive than compact particles, their mass-to-area ratios $m/A$ and Stokes numbers St can remain smaller, allowing the fractal models to retain much more mass in the region $\gtrsim 20$ AU over longer periods. This occurs because the aggregates further out in the disk have not yet compacted enough to have Stokes numbers comparable to compact particles. This is also reflected in panel (f) where aggregate ${\rm{St}}$ are larger inside 20 AU, but drop significantly below the ${\rm{St}}$ for compact particles outside this distance. The cold H$_2$O ice models (grey curves, panels b and f)  show this behavior only over a narrower region that corresponds to a modest range in temperature ($\Delta T \sim 20$ K), outside of which $Q_{\rm{f}}$ drops to the value for silicates so ${\rm{St}}$ decreases. As a result, $\Sigma_{\rm{solids}}$ remains significantly higher for the cold ice models than the fiducial models since the compact particles and aggregates have lower ${\rm{St}}$ and thus much longer drift times. In fact, in these models some material is being advected outwards due to the stronger gas coupling. Gas coupling also explains what is seen in the models for $\alpha_{\rm{t}} = 10^{-2}$ which also show no corresponding drop in $Z$ - because solid material is being even more strongly advected outwards in this case by the more strongly viscously evolving gas %\sout{However, we note that} 
The stronger gas motions for $\alpha_{\rm{t}} = 10^{-2}$  dominate because the ${\rm{St}}$ values for the highest turbulent intensity case are still small, even though larger than the cold ice, lower $\alpha_{\rm{t}}$ models because the nebula gas density is much lower. We refer the reader to \citetalias{Est21} for detailed discussion of the evolution of the particle mass distribution. 
 
In panels (c)-(e) of Figure \ref{fig:diskevol} we plot the midplane disk temperatures $T$, headwind parameter  %pressure gradient 
$\beta$, and the Rosseland mean opacity $\kappa_{\rm{R}}$. The headwind parameter %pressure gradient %\textcolor{red}{put this earlier as noted: $\beta \propto (c/v_{\rm{K}})^2$ 
(Eq. \ref{equ:pgrad}, Sec. \ref{sec:dustvapevol}; see also \citetalias{Est21}) is primarily what drives the inward radial drift of particles ($\propto \beta v_{\rm{K}}$) relative to the gas\footnote{Incidentally, at no point in these simulations does $\beta <0$, so no appreciable pressure bumps develop that could trap particles.}. Variations in these plots can largely be attributed to the evolution of the particle mass distribution and vice versa. In panel c, locations in the disk where the temperature flattens are just interior to EFs. The most refractory EFs have long ago evolved inwards off the grid, but the EFs for FeS, organics and water ice are still present. The CO$_2$ snowline is also present at $\sim 50$ AU; it is not visible in the temperature, but is visible as a peak or a kink in the surface density profile seen in panel b. The temperature profile for a MMSN ($T\propto R^{-1/2}$, \citealt{Hay81}) is given by the blue line. Variations in the Rosseland mean opacity largely mirror those of the solids surface density profile. We do not show the Planck mean opacity $\kappa_{\rm{P}}$, but these appear largely similar, except in the innermost disk and in the outer disk (mostly beyond $100$ AU) where the optical depth is low \citepalias[see Fig. 5,][for a comparison of these opacities for the fiducial model]{Est21}. It is interesting to note that the model for $\alpha_{\rm{t}}=10^{-2}$ begins with the highest initial temperature, while $10^{-4}$ begins with the lowest \citepalias[e.g., see Fig. 1,][]{Est16}, but it is the model with $\alpha_{\rm{t}} = 10^{-3}$ that remains the hottest after 0.5 Myr. As discussed in \citetalias{Est21}, models with $\alpha_{\rm{t}}=10^{-3}$ tend to be the ``sweet spot'' for the most optimal combination of particle growth and retention of solids in these disk models.

\subsection{Evolution of Solids and Vapors}
\label{sec:evosolvap}

\begin{figure}
\includegraphics[width=1.0\textwidth]{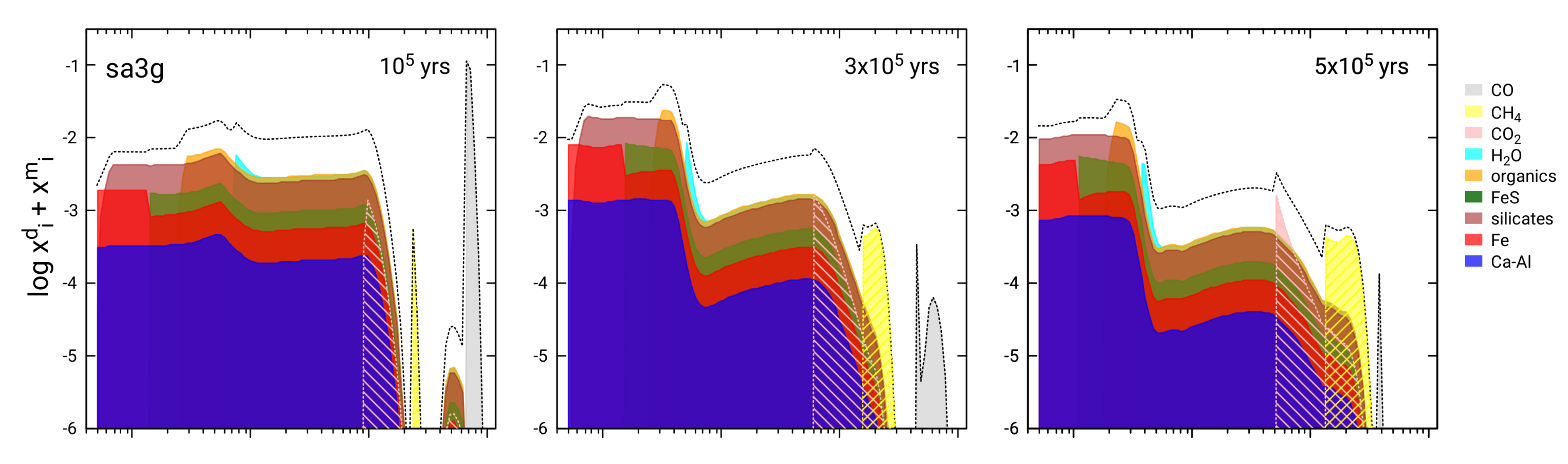}
\includegraphics[width=1.0\textwidth]{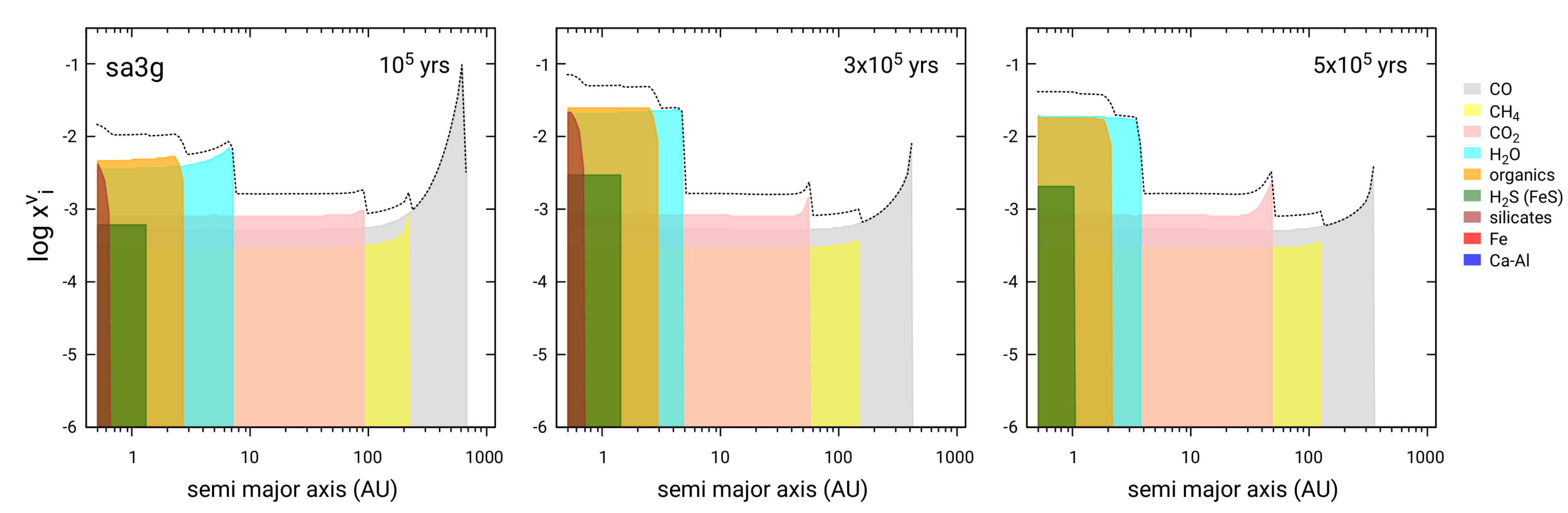}
 
\caption{The fiducial case ($\alpha_{\rm{t}}=10^{-3}$), for the compact particle growth simulation (sa3g, Table \ref{tab:models}), showing the evolution of solid and vapor fractions of all species (Table \ref{tab:species}) at 0.1, 0.3 and 0.5 Myrs. The location of EFs of the associated volatile species are characterized by enhancements in vapor inside (lower panels), and enhancements in solids outside (upper panels) these locations that can buildup with time. For example, the water snowline (cyan), which migrates from $\sim 7$ to $\sim 4$ AU over the simulation shown, reaches peak enhancement around 0.3 Myr. A ``tarline'' (orange) of organics around $\sim 2$ AU also peaks around the same time. The black dotted curves show the total of all species shown at the right (see Table \ref{tab:species}), and for the top panels, also represents the instantaneous metallicity $Z$ (sum of all solid species). The enhancement in $Z$ in the inner disk, and depletion in the outer disk, is due to particle radial drift inwards across the water snowline. The dotted colored curves and hatched regions in the upper panels trace the corresponding supervolatile solid fractions where those fractions would otherwise be obscured. See section \ref{sec:fiducial} for discussion.  
\label{fig:sa3g}}
\end{figure}

In this section we look at the evolution of the solid and vapor fractions of the individual species listed in Table \ref{tab:species} for all the simulations listed in Table \ref{tab:models}.  As a simplification in this work, we treat sublimation and condensation at EFs as completely reversible - which for our ``organic" species in particular is not realistic (see Sec. \ref{sec:discuss}). We will refine this treatment in a future paper. Our main goal herein is to study the radial and  temporal variation of solid and vapor composition for the different models due to the effects of advection, diffusion and radial drift. 

\subsubsection{Fiducial Model}
\label{sec:fiducial}

In Figure \ref{fig:sa3g} we first examine the compositional evolution for a compact particle growth model with our fiducial turbulent intensity  $\alpha_{\rm{t}}=10^{-3}$, also assuming a gaussian particle collision velocity pdf (case sa3g, see Paper II). Plotted in the top set of panels are the log of the solids mass fractions $x^{\rm{d}}_i + x^{\rm{m}}_i$ of the various species $i$, at three different times, 0.1, 0.3 and 0.5 Myr. The bottom panels are the corresponding vapor fractions $x^{\rm{v}}_i$. Though locations of EFs can be seen in both sets of panels, they are the most easily seen in the vapor. In both sets of panels, the black dotted curves are the total fractions of solids or vapors, respectively, which for the solids also represents the locally varying metallicity $Z$ (Sec. \ref{sec:dustvapevol}). 

{\it Mass Loss:} The first thing to note is the systematic depletion of solid material outside the snowline, that occurs over time due to inward radial drift of material. As a result the inner disk regions become enhanced with solids while the outer disk becomes depleted. The depletion is especially rapid {\it slightly} beyond the water ice EF (cyan) because the  mass-dominant particles here have the largest Stokes numbers ($\sim 0.02-0.03$) and thus the fastest radial drift \citepalias[see Fig. \ref{fig:diskevol} panel f, and][]{Est21}. Recall that for the  models shown, particles and aggregates can grow much larger everywhere beyond the water snowline  due to the stickiness of water ice (Sec. \ref{sec:dustvapevol}). The particle sizes that correspond to these Stokes numbers are $\sim 15$ cm for the compact particle case, and $\sim 27$ m for the fractal case (see Appendix A, \citetalias{Est21}). As a result, the local $Z$ drops by about an order of magnitude from the initial value in only $3\times 10^5$ years.  In the inner disk, the solids enhancement appears to peak around 0.3 Myr with $Z\sim 0.03-0.04$ between $3-4$ AU, primarily due to enhancement of solids outside the organics EF (orange). However, because the Stokes numbers of the mass dominant particles are small in the inner disk, with ${\rm{St}} \lesssim 0.001-0.01$, the local solids-to-gas {\it volume density ratio} $\epsilon = \rho_{\rm{solids}}/\rho \lesssim 0.1$ so streaming instability is precluded even for this $Z$  \citepalias[see][]{Est21}.  In any case, the magnitude of this narrow peak in $Z$ is probably an overestimate because sublimation-condensation of organics is likely not completely reversible as we assume here (see Sec. \ref{sec:cavfwk}, and \citetalias{Est21} for more discussion).
%\citep[see][]{Sen21}. 
On the other hand, the sublimation and condensation of troilite (FeS, green) is likely reversible \citep{Arm_1960}, but possibly only down to temperatures of $\sim 500$ K \citep[see, e.g.,][]{Pas05}.  When sublimated, FeS transitions to H$_2$S and refractory iron in the fraction 56/88 \citepalias[see e.g.,][]{Est16}. This explains the sharp increase in refractory iron (red) around $\sim 1$ AU, inside the FeS EF. For this work we implicitly assume condensation back to FeS outside the troilite EF if refractory Fe is available there (however see Sec. \ref{sec:cavfwk}).

{\it Evaporation Fronts:} In the outer disk, enhancements in supervolatiles CO (grey), CO$_2$ (pink) and CH$_4$ (yellow) can be seen outside their respective EFs, which evolve inwards as the disk cools (mostly due to decreasing stellar luminosity). A large enhancement spike in solid CO is seen at 700AU after 0.1 Myr, while all other solid species' fractions have systematically decreased in magnitude. This enhancement is not due to inward radial drift of particles, because early on in this region of the disk where the gas density is so low, even monomers have very large Stokes numbers and are effectively immobile. Rather, this enhancement is due to the nebula gas advecting outward as the gas disk spreads. In our models, the initial gas surface density profile drops off sharply outside of $\sim 100$ AU (see Fig. \ref{fig:diskevol}, panel a) so that even small changes can make a significant difference in the local solid fractions in the outermost parts of the protoplanetary disk. Even though the nebula gas density remains very low, it has increased by orders of magnitude over the initial value, resulting in decreased instantaneous $Z$ of all species except CO. %\textcolor{red}{A bit confused here; doesn't the expanding gas carry with it a mostly cosmic complement of the SVs in the vapor or sometimes fine dust grains? so how can it decrease the local fraction, as if condensible-free H2/He were moving out alone? Maybe (related to the following sentence) it is decreasing the $Z$ because the St of all the solid or even fractal particles already condensed at smaller radii is too large to be advected by the gas, so $Z_i$ for those species decreases. Need to clarify. }
The CO is enhanced because vapor phase CO {\it is} advecting with the nebula gas and condenses outside the CO EF %in a snowstorm - still not a lot of mass though} 
at a rate fast enough to produce the relatively large spike in solid CO fraction. The magnitude of the spike eventually decreases over time once
%is due to both (a) solid CO ice sublimation and diffusion back across the EF, and (b) the condensation, sublimation and recondensation of CO advected outwards with the nebula gas. However, these peaks shrink with time for two reasons: first, because the solid CO ice particles (as well as all other species) radially drift inward due to the strong pressure gradient (the disk edge drops off quite steeply, even after 0.5 Myr, see Fig. \ref{fig:diskevol}, panel a), but second and mostly 
the advected nebula gas begins to increase significantly enough beyond the CO EF as the disk continues to spread outwards. %\textcolor{red}{as mentioned above, this would require CO-free expanding gas though.}  
%This effect dominates evolution because, outside of $\sim 200-400$ AU, the Stokes numbers are quite large, even for very small particles, \textcolor{red}{ah so now we get the answer, maybe get to this sooner} so particles can effectively remain immobile until the gas density builds up (allowing their ${\rm{St}}$ to decrease). 
The particle Stokes numbers in this outermost region then decrease, to unity or less, at
%Within this radial region, ${\rm{St}}$ \textcolor{red}{then decreases to} $\sim 1$ at 
which point the region can be vacated due to rapid radial drift (e.g., see top left panel of Fig. \ref{fig:a4g}).

The methane and CO$_2$ EFs lie inside $\sim 200$ AU and thus exhibit a different behavior because the particle Stokes numbers are $\ll 1$. Both EFs have solids enhancements that extend outwards over many AU, with the CH$_4$ enhancement being especially extended. This is because (like the case for CO) each region is enhanced already by the outward diffusion of sublimated vapor and subsequent recondensation outside their respective EF, as well as the outward advection of the nebula gas, {\it but} here the nebula gas can effectively advect small particles with it as well. The peak for CH$_4$ is broader because the Stokes numbers of the mass-dominant particles are about an order of magnitude smaller than they are at the CO$_2$ EF, {\it and} the gas advection velocity $v_{\rm{g}}$  is increasing sharply in this region and is higher at the methane EF. Thus, small-St particles are advected further at the methane EF. %\textcolor{red}{rewrote some of previous sentence for clarity and brevity.}
% Their St >> 1 when the gas density is very small, may explain why others have had a cutoff. Would these particles even be there in the first place? Why not? Left behind somehow as the cloud collapsed?
 
 \begin{figure}
\includegraphics[width=1.0\textwidth]{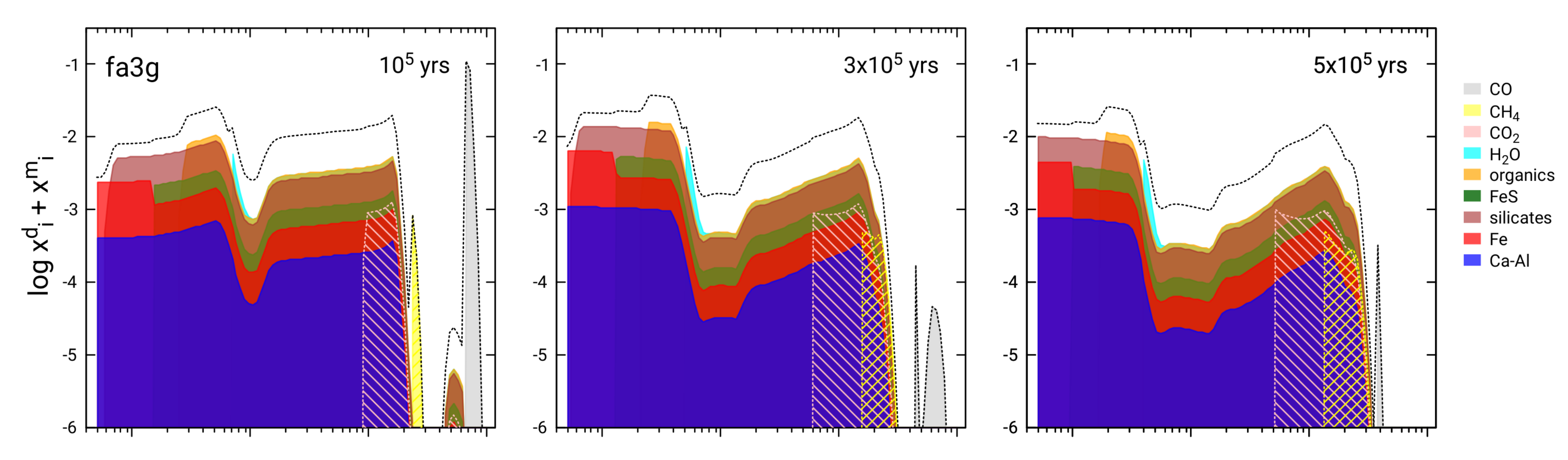}
\includegraphics[width=1.0\textwidth]{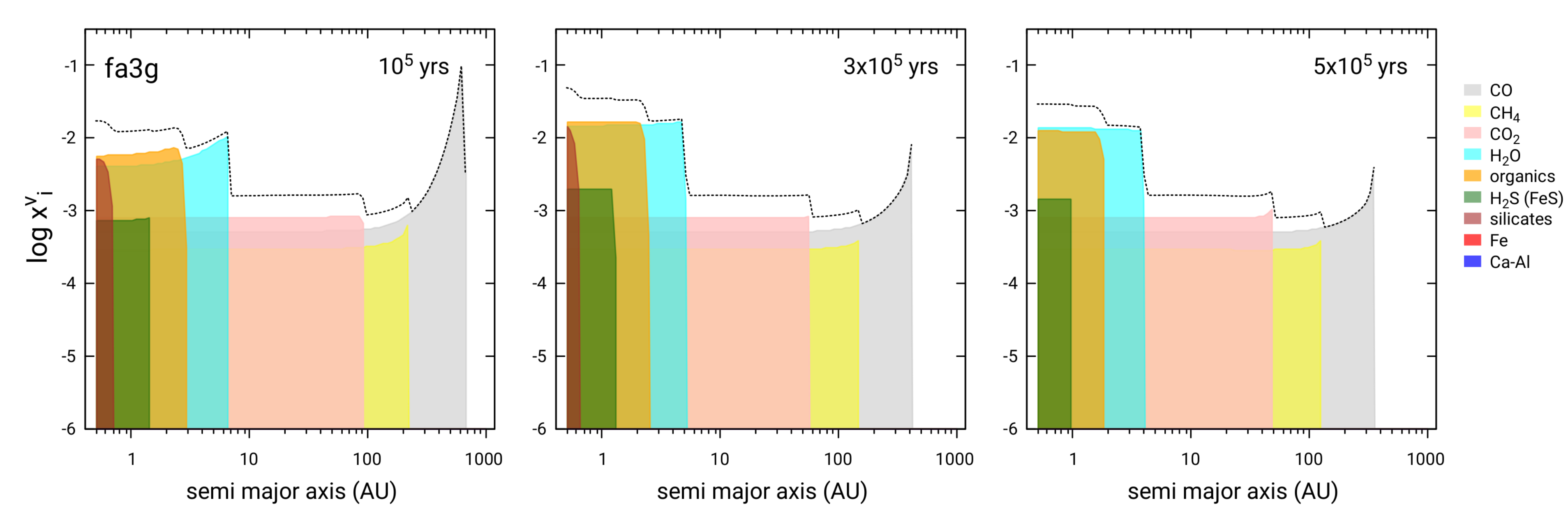}
 
\caption{Fractal aggregate growth simulation with the fiducial  $\alpha_{\rm{t}}=10^{-3}$ (fa3g) at 0.1, 0.3 and 0.5 Myrs, showing evolution of solid and vapor fractions of all species, for comparison with figure \ref{fig:sa3g}. The dotted colored curves in the upper panels are the corresponding supervolatiles solid fractions. As before, for the top panels, the black dotted curve represents $Z$. A main difference between this fractal case and the corresponding compact particle growth case in figure \ref{fig:sa3g} is that significantly more mass remains outside of $\sim 20$ AU. This is because fractal aggregates remain fluffier there, so their radial drift speeds are lower than the compact growth case, which is reflected in the difference in St$_{\rm{max}}$ values (orange curves, Fig. \ref{fig:diskevol}, panel f) between the two models outside this radial location.
\label{fig:fa3g}}
\end{figure}

The bottom panels of Fig. \ref{fig:sa3g} indicate that there are seven EFs still on the computational grid after 0.1 Myr (all 9 are present at $t=0$), so for the times shown, both iron and Ca-Al (``CAIs'', blue) are solid everywhere in the disk. The silicate EF (brown) persists for quite some time, but has  evolved inwards off our inner grid boundary at 0.5AU by 0.5 Myr. Because the location where the nebula gas velocity switches from outwards to inwards occurs at $\sim 20$ AU, the enhancements in solids outside the various EFs in the inner disk tend to be due purely to diffusion for these models, while similar EF enhancements outside 20AU  are due partly to advection. It is noteworthy that the enhancements in the metallicity of the {\it gas} can also become substantial (5-8\%). This could systematically increase the local gas molecular weight, affecting the pressure gradient (and viscosity) through both the gas density and the local sound speed \citep[see, e.g.][]{Cha21}. A feedback effect from the increasing molecular weight of the gas might possibly lead to a pressure bump. This becomes more of an issue for lower $\alpha_{\rm{t}}$. We will consider this effect more carefully in future work.

\begin{figure}
\centering
\includegraphics[width=0.4\textwidth]{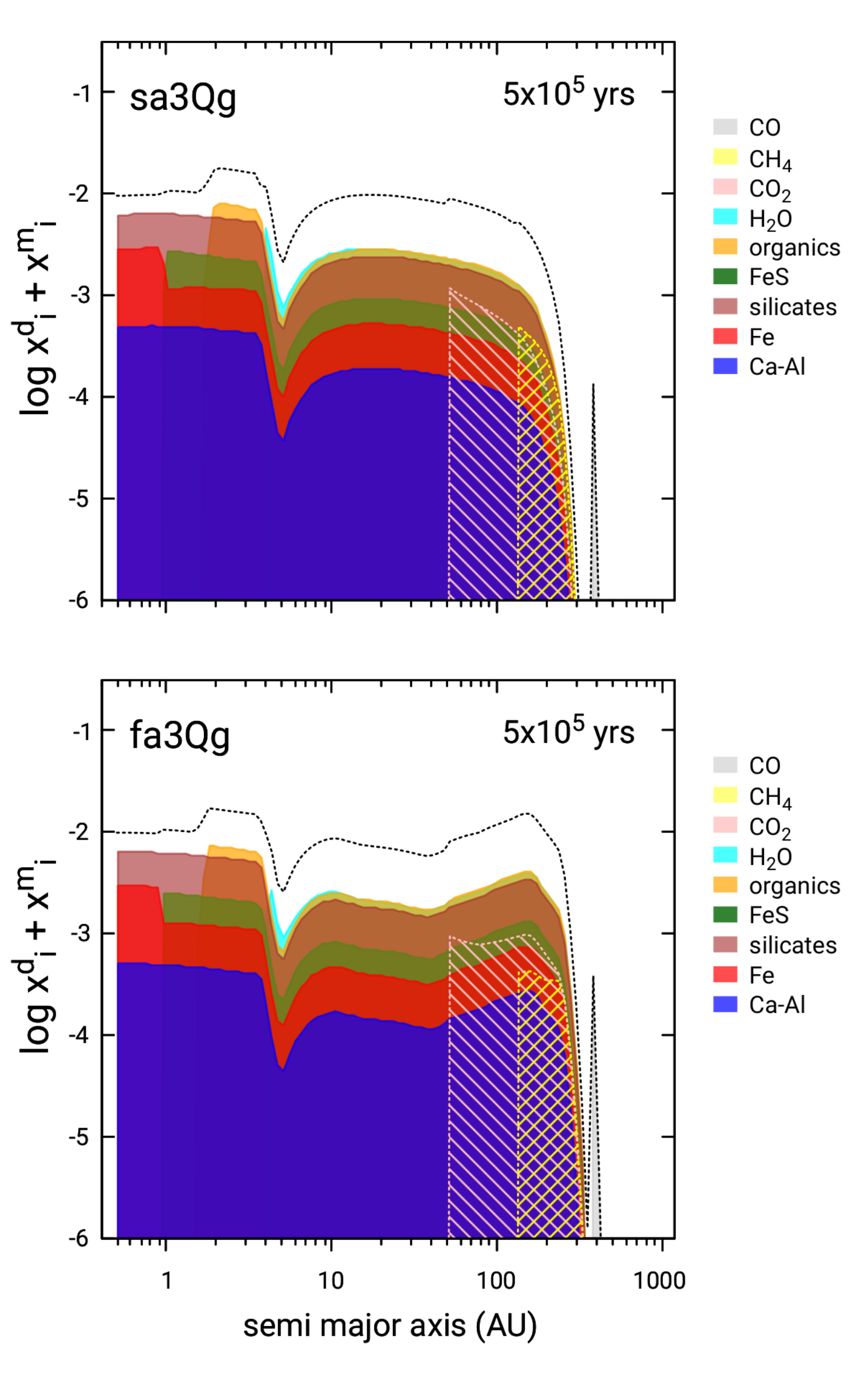}
\caption{``Cold H$_2$O Ice" simulations for solids, showing  both  fractal aggregate (fa3Qg) and compact particle (sa3Qg) cases, after 0.5 Myr in the case where water ice is sticky only over a limited temperature range about the snowline  \citep{MW19}. These models also assume the fiducial  $\alpha_{\rm{t}}=10^{-3}$. As before, the black dotted curve in the upper panel represents $Z$. 
\label{fig:a3Qg}}
\end{figure} 

{\it Fractal aggregate vs. compact particle growth:} In Figure \ref{fig:fa3g}, we show a fractal aggregate growth model (fa3g) for the fiducial $\alpha_{\rm{t}}=10^{-3}$ for comparison with the  compact particle growth simulations of Figure \ref{fig:sa3g}. Many of the trends described for the compact growth case are also seen here. For example, the sharp decrease in total solids abundance %to $Z\sim 10^{-3}$ 
outside the water ice EF ($\sim 6-15$ AU) still occurs, with the solids surface density profiles overall being very similar (Figure \ref{fig:diskevol}). Like in model sa3g, the metallicity in the region from the H$_2$O EF out to $\sim 15$ AU (dotted black lines in top panels of Fig. \ref{fig:sa3g} and \ref{fig:fa3g}) has dropped to $Z\gtrsim 0.001$, %\sout{though the fractal ${\rm{St}}$ are slightly higher than} \textcolor{red}{
because the St in this region are similar for sa3g and fa3g %for the solid particle case 
($\sim 0.03$; see Fig. \ref{fig:diskevol}, panel f)\footnote{It should be understood that growth to large ${\rm{St}}$ and decreasing $Z$ occur concomitantly. That is, the local $Z$ is low {\it because} the local ${\rm{St}}$  is large, and $Z$ {\it and} ${\rm{St}}$ are never both large enough at the same time and place that conditions for the Streaming Instability can be satisfied \citepalias[see][and also \citealt{Umu20}]{Est16,Est21}.}. %\textcolor{red}{this sentence was unclear. The St look about the same inside of 20AU, and then the fractal St's drop below those for the solids all the way to 200AU. }
The main difference is that now beyond $\sim 20$ AU much more material is retained because the fluffy aggregates have Stokes numbers $5-10$ times smaller than the compact particle case (orange curves, Fig. \ref{fig:diskevol}, panel f), though the fractal masses are several orders of magnitude larger than their compact counterparts (see Fig. 2, \citetalias{Est21}). Again, this is because aggregates have not yet compacted enough, so their St remain lower. The aggregates thus have much slower inward radial drifts, and a fraction of them can even be advected outwards. Indeed, where we saw much of the solids being depleted in the region $\gtrsim 40-100$ AU in Fig. \ref{fig:sa3g} leaving a clear enhancement in narrow supervolatile bands of CO$_2$ and CH$_4$ solids, the outer edge of this region remains populated with particles in the fractal case and has actually moved outwards from $\sim 200$ to $\sim 300$ AU over 0.5 Myr. The supervolatile bands are still present for the fractal case, but are very much muted (see Sec. \ref{sec:partcomp}). The differences can also be seen in the $\Sigma_{\rm{solids}}$ profiles (panel b, Fig. \ref{fig:diskevol}). %The fractal porous aggregates are many orders of magnitude more massive than their solid particle counterparts \textcolor{red}{particle masses are not shown in figure 1, but in paper I, do you mean the $\Sigma_{\rm{solids}}$ values?}, which have fairly similar St  \citepalias[as is generally true; Figure \ref{fig:diskevol}f, and][]{Est21}. As mentioned before, the large area per unit mass of the fractal aggregates keeps their ${\rm{St}}$ low at this juncture in the evolution. 
Eventually though, we expect even the fractal aggregates outside $\sim 20$ AU will continue to compact and drift into the inner disk. The vapor fractions are similar in both models with the exception that the enhancements inside the supervolatile EFs are somewhat reduced.

{\it ``Cold (non-sticky) H$_2$O ice":} In Figure \ref{fig:a3Qg} we show  simulations for compact particle (sa3Qg) and fractal aggregate (fa3Qg) growth simulations for $\alpha_{\rm{t}}=10^{-3}$, which differ from the fiducial model in that we here assume that water ice is only ``sticky'' in a limited temperature range near the snowline  \citep{MW19}. We implement this in our simulations by only using the ``sticky ice" $Q_{\rm{f}} = 10^6$ (see Table \ref{tab:models}) within 20 K of the water ice condensation temperature, and adopting the lower (silicate) value of $Q_{\rm{f}}$ at lower temperatures. In Fig. \ref{fig:a3Qg} we only show the solids fractions for the ``cold ice" models at 0.5 Myr. One finds that for both fractal and compact particles, the depletion outside the snowline has been significantly reduced, because particles and aggregates cannot grow to nearly as large St further out in the disk, decreasing the mass influx of material. The region near the H$_2$O EF in which $Q_{\rm{f}}$ and  the fragmentation threshold remain high still has larger particles, with ${\rm{St}}$ similar at their peak to the fiducial case (Fig. \ref{fig:diskevol}, panel f), but dropping off  by nearly 2 orders of magnitude from there outwards.  Enough material drifts in to prevent as large a drop in $Z$ as seen in the fiducial model, but  because less material drifts across the water ice EF overall, the enhancements in the inner disk are also smaller. The fractal case fa3Qg shows slightly more retention of material in the outermost portions of the disk than the corresponding compact particle case sa3Qg, but the latter retains much more than the fiducial compact particle growth case sa3g due to the smaller ${\rm{St}}$ (Fig. \ref{fig:diskevol}). %\textcolor{red}{I don't know what the following sentence means or refers to, or how it connects to what you just said..} As a result, the buildup of supervolatiles like CH$_4$ and CO$_2$ are similar to the  fractal models. 

\subsubsection{Variation with Turbulence Parameter $\alpha_{\rm t}$}
\label{sec:varturb}

\begin{figure}
\includegraphics[width=1.0\textwidth]{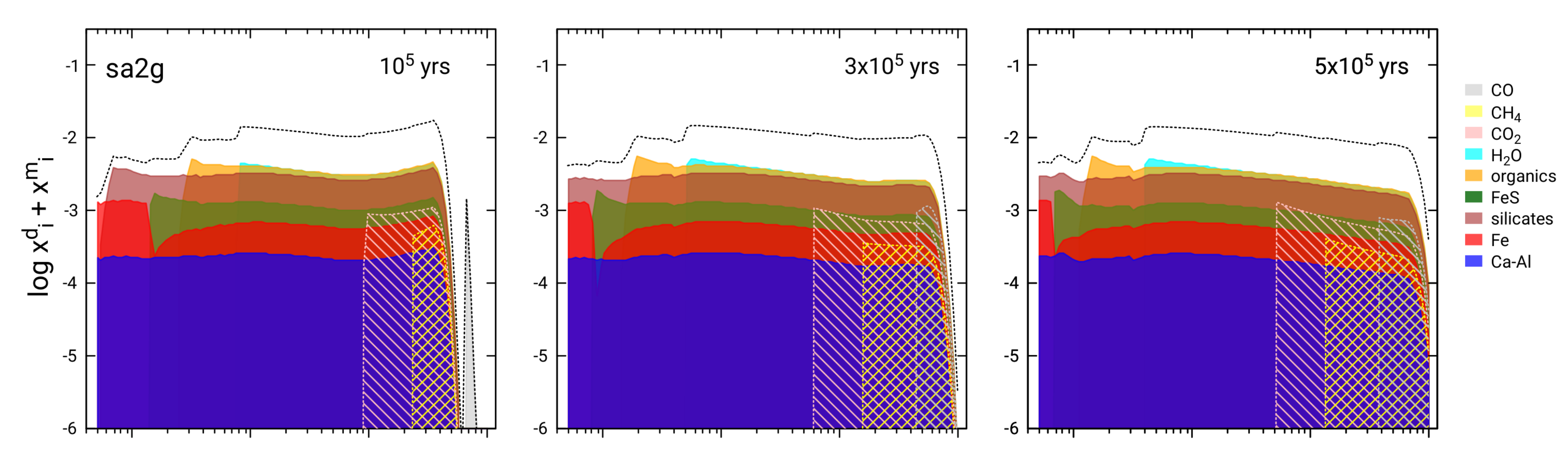}
\includegraphics[width=1.0\textwidth]{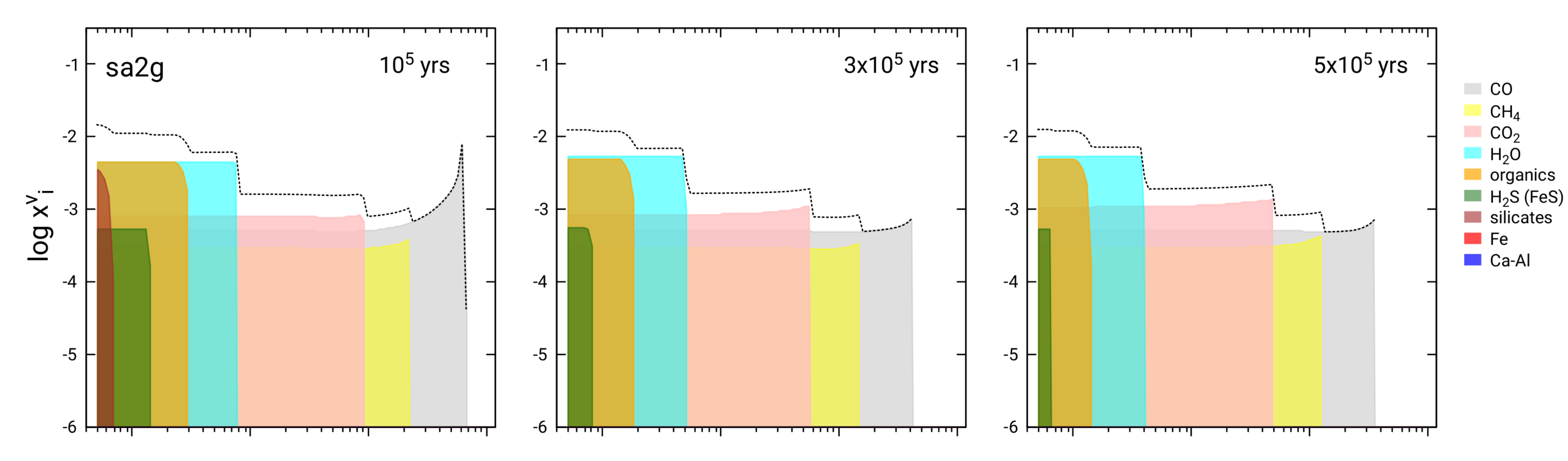}
\includegraphics[width=1.0\textwidth]{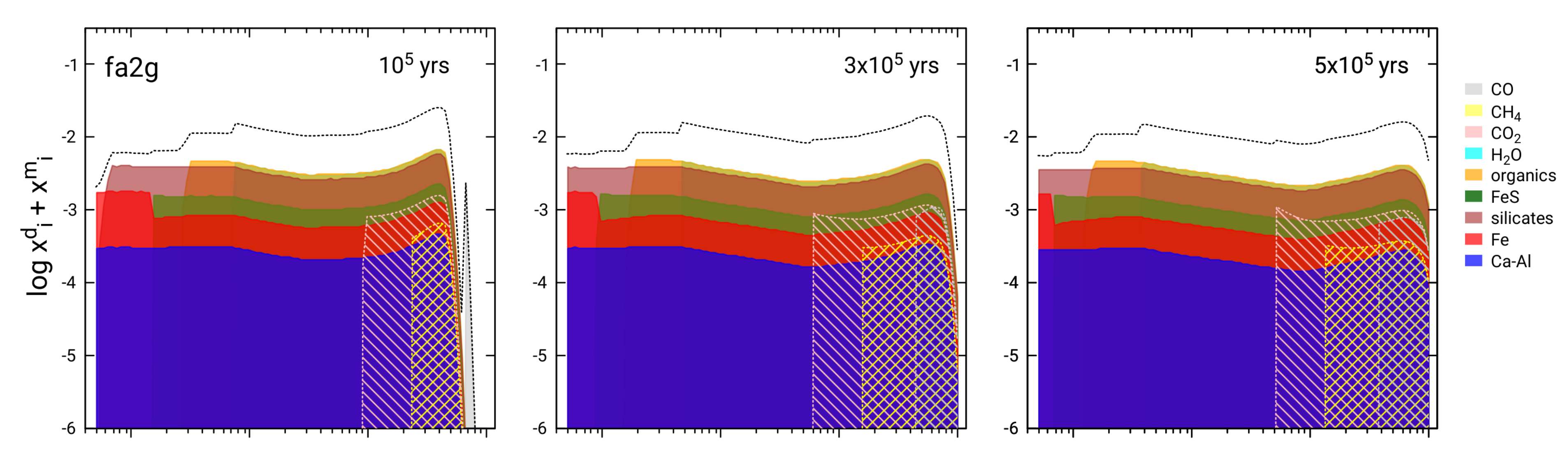}
\includegraphics[width=1.0\textwidth]{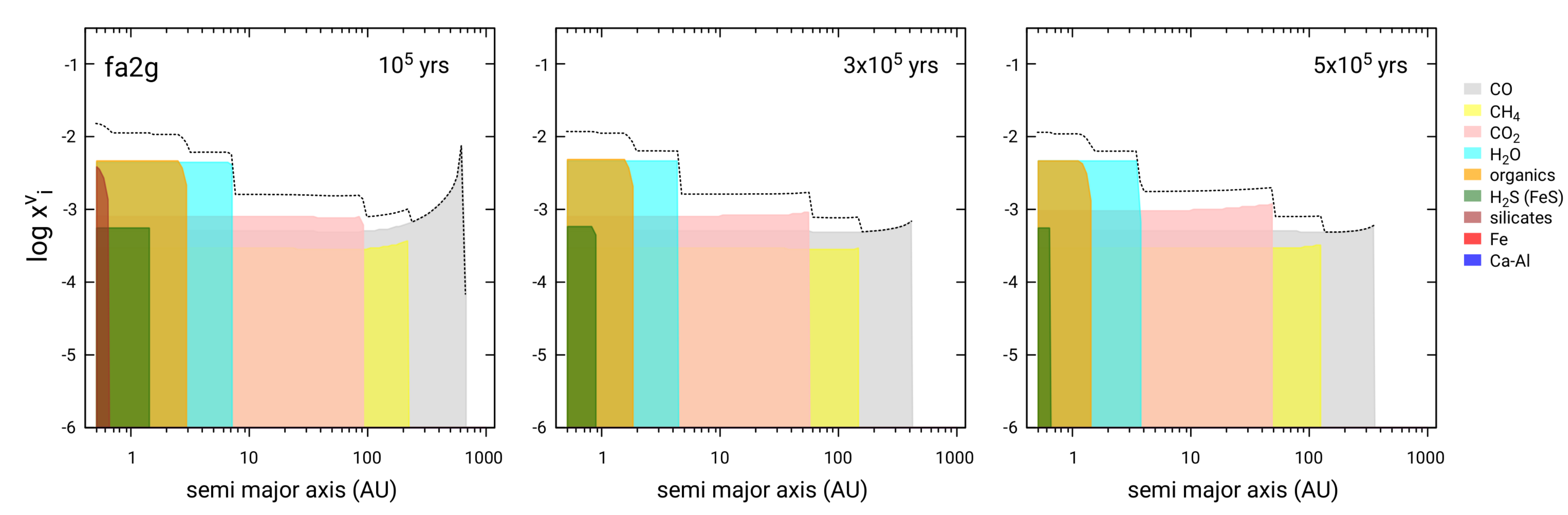}
\caption{Simulations with $\alpha_{\rm{t}}=10^{-2}$ of the evolution of solid and vapor fractions of all species for the compact particle (sa2g) and fractal aggregate growth (fa2g), at 0.1, 0.3 and 0.5 Myrs. The black dotted curves show the total of all species, and for the 1st and 3rd row of panels represents the instantaneous metallicity $Z$. The dotted colored curves in the 1st and 3rd rows are the corresponding supervolatile solid fractions. The similarity between sa2g and fa2g (certainly for the vapor phases) imply that the differences between fractal aggregate and compact particle growth models begins to disappear with increasing $\alpha_{\rm{t}}$.
\label{fig:a2g}}
\end{figure}

{\it Larger $\alpha_{\rm t}$ {\rm ($10^{-2}$)}:}
Simulations where we increase the turbulence parameter to $\alpha_{\rm{t}}=10^{-2}$ are shown in Figure \ref{fig:a2g}.  Because relative velocities between compact or aggregate particles increase with $\alpha_{\rm t}$, the fragmentation sizes and Stokes numbers are considerably smaller (at least initially) so that they remain more well coupled to the nebula gas and less affected by radial drift. As a result, the enhancements of solids and vapors at EFs are more muted overall compared to the fiducial models, and less material is lost from the outer disk. There is essentially no difference seen between compact and fractal growth in the evolution of the vapor species (2nd and 4th rows) over the course of the simulation.

{\it New result on Fe and FeS:} However, some interesting behavior is seen for the particles in the inner disk. In the compact growth model (1st row), the condensed FeS outside its EF appears to have the strongest enhancement - similar to or even larger than seen at the organics and water EFs. This FeS-enrichment comes at the expense of a dramatic depletion in iron metal near $0.9$ AU.  This interesting effect, which actually begins around 0.2 Myr (not shown) in both fractal and compact growth models, occurs because the quick diffusion of a substantial amount of H$_2$S across the FeS EF is reacting with inward drifting, native Fe metal to produce a large spike in condensed FeS, but the Fe metal is not replaced as quickly by inward radial drift of material from further out. By 0.5 Myr and beyond though, enough Fe dust is drifting in to begin to erase the Fe-depletion signature. This localized Fe-depletion is not seen in the $\alpha_{\rm t}=10^{-3}$ fiducial model %\sout{(though see below)} 
where the enhancement in FeS is absent, and instead  the refractory iron is enhanced interior to the troilite EF. This Fe-depletion disappears sooner in the fractal aggregate growth model (mostly absent by 0.3 Myr) for  this large $\alpha_{\rm t}$ (3rd row of Fig. \ref{fig:a2g}) because, due to slower radial drift (smaller St in the inner disk overall), the enhancement of H$_2$S, and its feedback across the EF, does not build up sufficiently to cause a depletion in refractory iron for nearly as long. We discuss this %new effect 
further in section \ref{sec:partcomp}.

{\it Global evolution with time for vapor and solids:} As pointed out above, there is no notable difference  between the compact and fractal cases in the vapor evolution for the $\alpha_{\rm t}=10^{-2}$ models, and at 0.1 Myr, the solids fractions in the outer disk regions beyond the snowline (1st and 3rd rows) also look quite similar. This changes gradually at later times. For both fractal and compact cases as the gas disk spreads outwards and the gas surface density increases significantly beyond $\sim 100$ AU (see panel a, Fig. \ref{fig:diskevol}), particles and aggregates are more easily advected outwards with the nebula gas, whose velocity $v_{\rm{g}}$ increases strongly beyond $\sim 20$ AU. Outside $\sim 90$ AU,   the ${\rm{St}}$ of the fractal aggregates are smaller than their compact counterparts and are thus more coupled to the gas, so they are transported outward in greater abundance. This explains the slight difference in $Z$ between the models at large distances at later times.

The rapid gas evolution for $\alpha_{\rm t}=10^{-2}$ also decreases the gas density everywhere inside of $\sim 100$ AU significantly, and  ${\rm{St}}$ generally increases (red curves in Fig. \ref{fig:diskevol}, panel f) to values comparable to or even larger than the lower $\alpha_{\rm{t}}$ models %, \textcolor{red}{do you mean, while St increases, it still remains small,  .. you rewrote a lot of this since the draft I read..} so it is mostly the strong advection driving material outwards. 
despite being smaller in mass in comparison \citepalias[{\it cf.} Figs. 2 and 8,][]{Est21}. %their smaller masses overall, t
The differences in particle mass between the compact and fractal cases is still a few orders of magnitude \citepalias[Fig. 8,][]{Est21}, but the Stokes numbers of the mass dominant fractal aggregates are roughly the same for the compact and fractal growth cases, between the snowline and 90AU. %At around $\sim 90$ AU, however, the ${\rm{St}}$ for the fractal aggregates and solid particles diverge (the former are smaller) which corresponds to a slight uptick in $Z$. 
We note that the much higher level of coupling between particles and the gas in both cases more or less maintains $Z$ close to its initial value across the disk. This is largely true even in the inner disk, though there are slight enhancements in the solids outside EFs out to the snowline (including even the silicates at 0.1 Myr) in the compact particle growth simulation. Overall it appears the distinction between compact particle and fractal aggregate models diminishes for increasing $\alpha_{\rm{t}}$.

\begin{figure}
\includegraphics[width=1.0\textwidth]{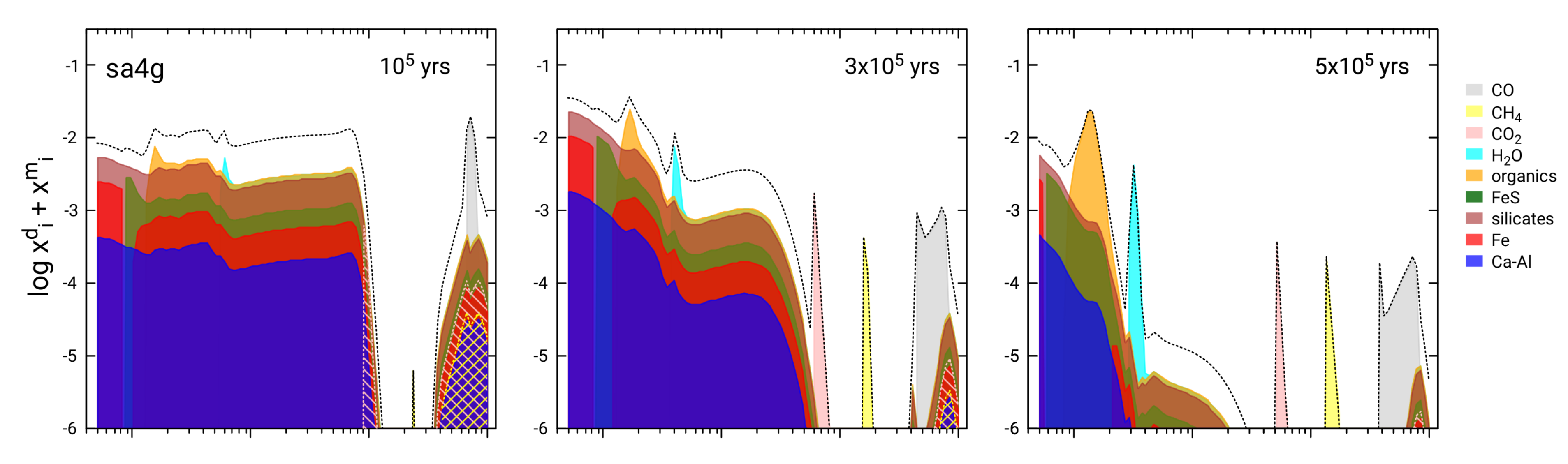}
\includegraphics[width=1.0\textwidth]{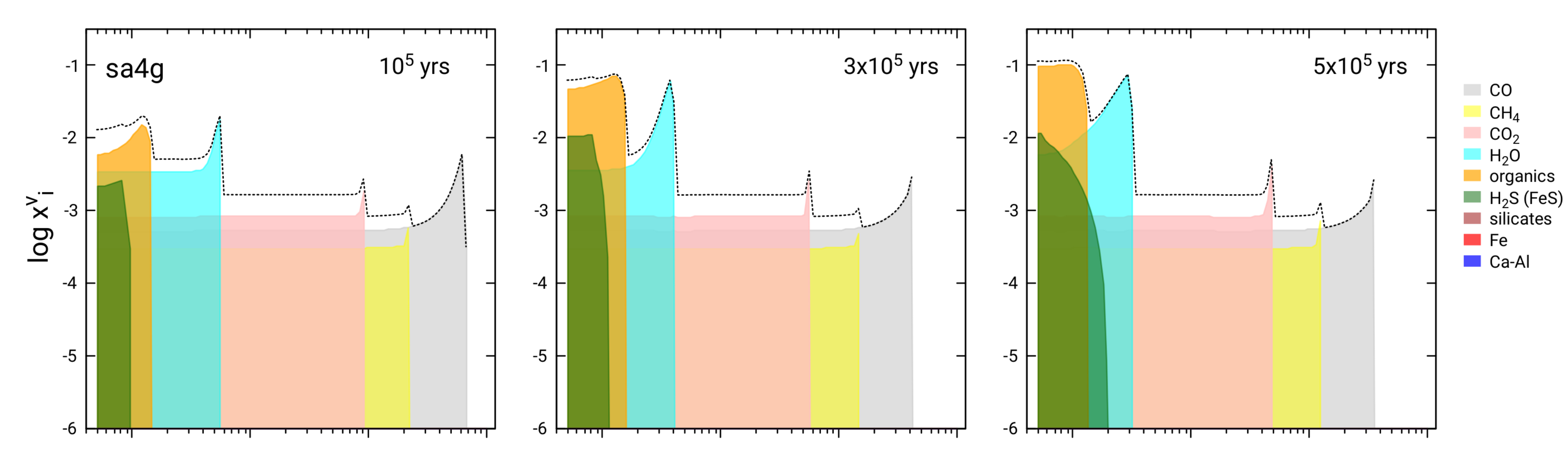}
\includegraphics[width=1.0\textwidth]{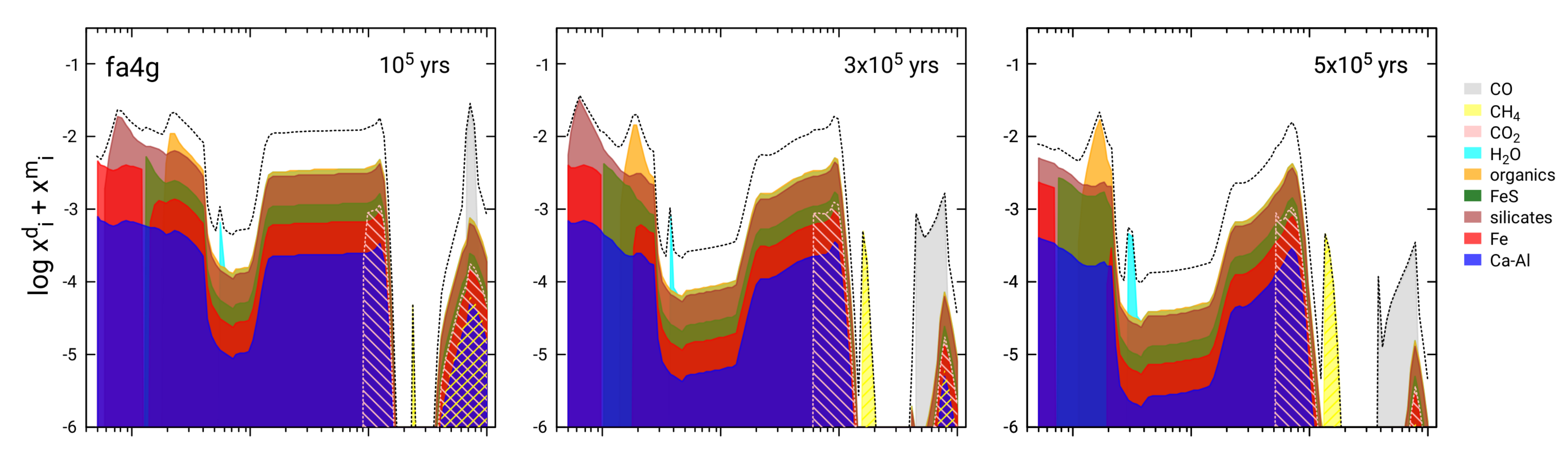}
\includegraphics[width=1.0\textwidth]{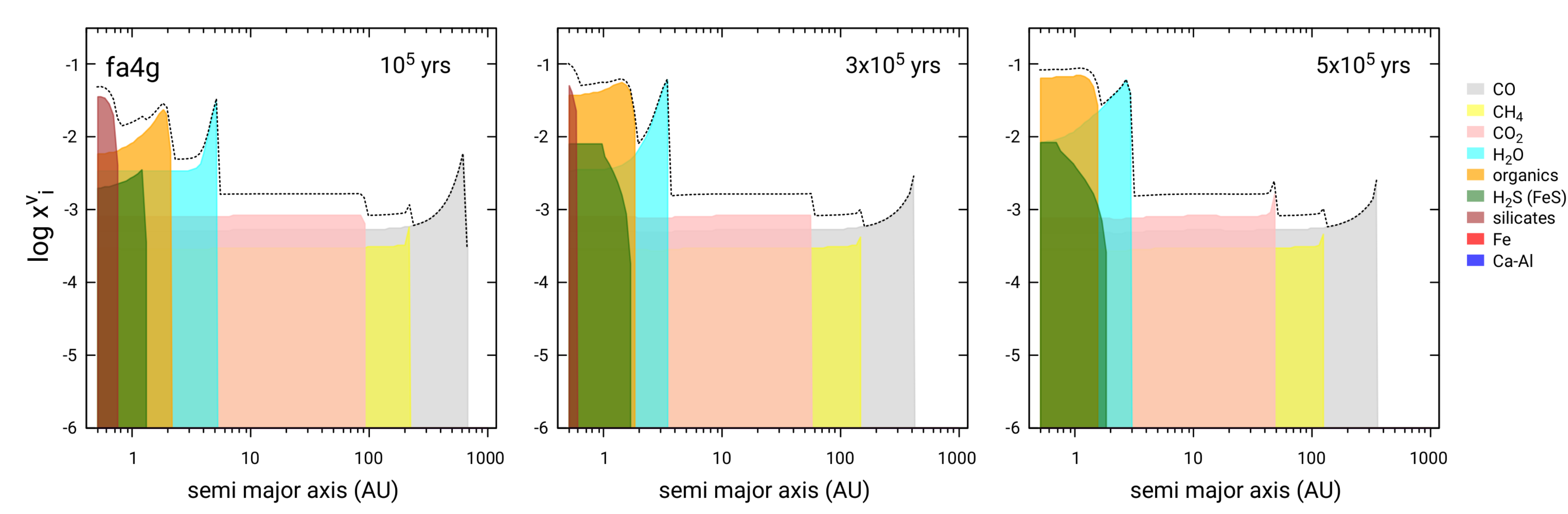}
\caption{Simulations with $\alpha_{\rm{t}}=10^{-4}$ of the evolution of solid and vapor fractions of all species for the compact particle (sa4g) and fractal aggregate  (fa4g) growth cases, at 0.1, 0.3 and 0.5 Myrs. The black dotted curves show the total of all species, and for the 1st and 3rd row of panels represents the instantaneous metallicity $Z$. This lower value of $\alpha_{\rm{t}}$ enhances the difference between fractal aggregate and compact growth models, especially outside the water snowline. While most of the mass in sa4g has drained into the inner disk by 0.5 Myr, significantly more material remains in fa4g, %\sout{even in the region between $\sim 5-10$ AU.} \textcolor{red}{
especially in the region outside 10AU. Another %\textcolor{red}{
interesting property of this $\alpha_{\rm{t}}$ is the higher vapor abundance inside the snowline, especially for fa4g, than for $\alpha_{\rm{t}}=10^{-2}$. This seems to be because  more material has drained into the inner disk,  evaporated there, and remains there because of the more sluggish overall disk evolution.
\label{fig:a4g}}
\end{figure}

{\it Smaller $\alpha_{\rm t}$:}  The change in overall behavior for smaller $\alpha_{\rm{t}}$  is much more dramatic, especially for compact particle growth, as can be seen for the case of $\alpha_{\rm{t}}=10^{-4}$ plotted in Figure \ref{fig:a4g}. The top panel (1st row), for the compact particle growth case, shows that the outer disk is much more quickly drained of solids to the inner disk regions compared to the fiducial $\alpha_{\rm{t}}=10^{-3}$ model for compact particles, and also that  material within the inner disk continues to be lost through the inner boundary and presumably to the star. The ${\rm{St}}$ values of the mass dominant particles for both fractal and compact growth models are still relatively small and similar to each other (Fig. \ref{fig:diskevol}), but now an order of magnitude larger than in the fiducial model (so is the gas surface density). Similar St between sa4g and fa4g also means the loss rate of particles is roughly the same for the compact and fractal aggregate growth cases (1st and 3rd rows) inside the snowline despite the aggregates being 2 orders of magnitude larger in mass than the compact ones \citepalias[see Fig. 8][]{Est21}. %, but this is not surprising since their ${\rm{St}}$ are roughly the same between sa4g and fa4g (Fig. \ref{fig:diskevol}). 
The exception is the extended knee inwards of the snowline from $\sim 2-3$ AU seen in Fig. \ref{fig:diskevol} (panel f). %\textcolor{red}{but wait... just above, you said at least the solid particles were larger than the fiducial ones. I have some confusion about what region, or what models, you are talking about here. } 
We note that this region inside the snowline %\textcolor{red}{.. this what? region? case? } 
corresponds to a sharp drop in the normalized pressure gradient, or headwind parameter $\beta$ (panel d), which would slow the radial drift rate in the region inside the snowline. For the fractal aggregates, the temperature (panel c) is  hotter and has a steeper gradient in this region (which corresponds to the organics EF). Coupled with the enhancement in organics, this drop in the headwind parameter  %pressure gradient drop 
allows aggregates that decreased in mass (due to loss of water) upon crossing the snowline to grow significantly larger again (compared to the compact particle case, see black curves Fig. \ref{fig:diskevol}, panel f) before drifting to the organics EF where considerable  material is evaporated, so that aggregate Stokes numbers and particle masses plummet. The spike in water ice outside the snow line is also lower, meaning that aggregates that cross the snowline are less ice rich compared to the compact particle case, so they lose less mass as they cross this EF. The same effect is not seen in model sa4g because the dip in $\beta$ is further inward and not as large in magnitude. 

Both fractal and compact particle growth cases see large enhancements in vapors inside, and in solids outside, the organics EF giving local (solids) $Z$ as high as $\sim 0.03-0.04$. The overall $Z$ curves inside the snow line look very similar in magnitude  for both models, but differences in their $\Sigma_{\rm{solids}}$ are notable (Fig. \ref{fig:diskevol}, panel b). The silicates EF also contributes to the local enhancement of $Z$, and persists longer in the fractal aggregate case. That is, the compact growth model cools much more quickly (see below). The enhancements in vapor (2nd and 4th rows) are considerably larger than in the fiducial model after 0.5 Myr, reaching values as high as $\sim 0.1$ and helping to produce the large enhancement in solid organics exterior to its EF. However, as noted previously, the solids fraction just outside the organics EF may not be realistic because the sublimated or decomposed CHON organics might break down into more volatile constituents such as CO or CH$_4$, rather than recondense as a CHON. This would not greatly affect the enhancement to the total vapor fraction, which  gets quite large in this region at least until these constituents have time to diffuse. As mentioned in Sec. \ref{sec:fiducial}, such large values for the vapor fraction further indicate that consideration of changes in the gas' molecular weight may  become important.

{\it Fe-FeS behavior:} The depletion of refractory iron  over a small radial range is even more dramatic in this   $\alpha_{\rm{t}}=10^{-4}$ case than it was in the case for $\alpha_{\rm{t}}=10^{-2}$, and unlike the latter case, the depletion is not temporary. For both the compact and fractal growth cases (1st and 3rd rows), the depletion of refractory iron begins very early ($< 0.1$ Myr) and the radial region over which this depletion grows increases with time as the excess diffusing H$_2$S gas gobbles up available Fe, maximizing the amount of troilite that can be sublimated again at the EF. The feedback creates so much excess H$_2$S that it diffuses further outward until it can react with more refractory iron further out in the disk. This process thus creates a situation in which troilite and H$_2$S coexist out to as far as $\sim 2$ AU in the compact growth case, and somewhat less in the fractal case (2nd and 4th rows), though with time we would expect the situation to extend further out, %\textbf{(though, see Sec. \ref{sec:cavfwk})}, 
as it did in the case with $\alpha_{\rm t}=10^{-2}$. Recall this effect is not seen in the fiducial model.  Although it is not clear why the $\alpha_{\rm{t}}=10^{-3}$ model does not show this effect, we suspect that it is because the inner disk remains hotter, longer than the lower and higher $\alpha_{\rm{t}}$ models (see Fig. \ref{fig:diskevol}, panel c). %\textcolor{red}{You should say something about why; you may in section 4.1..} 
Despite the more rapidly drifting compact and aggregate particles (due to their higher St overall), %\textcolor{red}{?? ..we have learned that more massive does not mean larger St.. it would be good to clarify this as meteoriticists will be very interested..}, 
Fe still cannot drift in fast enough to overcome this effect; presumably the effect could continue until most of the S from a range of radii is utilized (for more discussion see section \ref{sec:partcomp}).

Much like we saw in the fiducial case, the region outside the snowline is significantly depleted in solids, as growth there is  rapid due to the higher threshold $Q_{\rm{f}}$ for ice. However, depletion in this region is especially rapid and substantial in the fractal aggregate growth model (3rd row). Even after 0.1 Myr the region between $\sim 4-15$ AU sees a large drop in the local metallicity that is absent in the compact particle case at this juncture. The more rapid growth of fractal aggregates comes about in large part due to their larger cross sections \citepalias{Est21}. As aforementioned, particle Stokes numbers inside the snowline are similar between the compact and fractal models (with the exception of the knee region from $\sim 2-3$ AU), %\textcolor{red}{need to describe this knee thing better})
but outside the water ice EF fractal growth has led to mass-dominant aggregates with ${\rm{St}}$ about a factor of 5 larger than the compact particles. In fact, by this time, the aggregate masses are already as much as 6 orders of magnitude more massive than the compact particles in the same region \citepalias[see Fig. 9 of][]{Est21}. This leads to initially more rapid drift of solids into the inner disk regions and explains why, at the early time of 0.1 Myr, there is more enhancement in the inner disk for the fractal case. But the rapid depletion means that by later times the net inward flux of material is lower than in the compact case which steadily drains into the inner disk. This is because  in the fractal case, the mass-dominant aggregate ${\rm{St}}$ drops below that of the compact case outside of $\sim 15-20$ AU   by a factor $\sim 5-10$ (Fig. \ref{fig:diskevol}, panel f) so their drift rates are much slower, and considerable material remains there. On the other hand, rapid, continuous drift in model sa4g accounts for the much larger enhancement spike at the snowline in that model. By 0.5 Myr, the compact growth case has effectively lost most all of its material outside the snowline, while the fractal case retains a significant fraction.

{\it Supervolatile bands?} An effect of strong radial drift of disk material in the compact growth model case is the ``trapping'' of material that manifests as  bands of supervolatiles centered  at $\sim 55$AU (CO$_2$), $\sim 150$AU (CH$_4$) and beyond $\sim 400$AU (CO). The magnitudes of these spikes in $Z$  decrease over time, but for different reasons, as discussed for the fiducial model. We noted before that the  region between $\sim 200-400$ AU is quickly vacated\footnote{This specific radial range is not special but will vary with the choice of $R_0$. A larger (smaller) value of $R_0$ would move this location outwards (inwards). Indeed, a smaller $R_0$ may better correspond to an early stage disk subject to infall.}, even for small particles, because the Stokes numbers are already of order unity or larger there, so radial drift is  rapid. On the other hand, outside $\sim 400$ AU, Stokes numbers become increasingly large due to the sharply decreasing gas density so the particles and aggregates are not drifting much at all.  %\textcolor{red}{I think I scrubbed a mention of SI maybe being successful late in nebula evolution when the gas density decreases, because (as we discussed recently) it's usually found that as the gas density decreases the particle size just decreases in tandem to keep ST the same. If not, maybe you need to expand on that idea here. Is this just a minor fringe thing or is it important in planetesimal forming regions? CAN growth overcome drift? } 
The decrease in mass fraction there is solely due to the slow expansion of the nebula gas disk over time, which in this low $\alpha_{\rm{t}}$ case is too slow to diminish the CO peak after 0.5 Myr compared to the fiducial model. Also as before, the band of CH$_4$ is seen to be more extended in the fractal model due to the slower drift rates of the porous aggregates compared to the compact particles. It should be noted that these bands do not have a lot of mass, at least at this epoch, as can be gleaned from panel (b) of Fig. \ref{fig:diskevol}. For the compact growth model where a clear CO$_2$ band is seen, there is a peak surface density of $\sim 0.01$ g cm$^{-2}$, while for methane it is a hundred times smaller. The fractal model has a similar magnitude for CH$_4$, but at the CO$_2$ EF most species (Fig. \ref{fig:a4g}, 3rd row and panel) %\textcolor{red}{??????} 
are still present thanks to the low radial drift speeds of the porous, fluffy aggregates there (see also Fig. \ref{fig:partcomp}) so the fraction of CO$_2$ is 2 orders of magnitude larger. %\textcolor{red}{It seems the CO, CO$_2$, and CH$_4$ bands are all comparable, can you give numbers for $\Sigma_{\rm solids}$? } %\textcolor{red}{Need to say more about the $\Sigma$ and $\tau$ of these features. }

\begin{figure}
\centering
\includegraphics[width=1.0\textwidth]{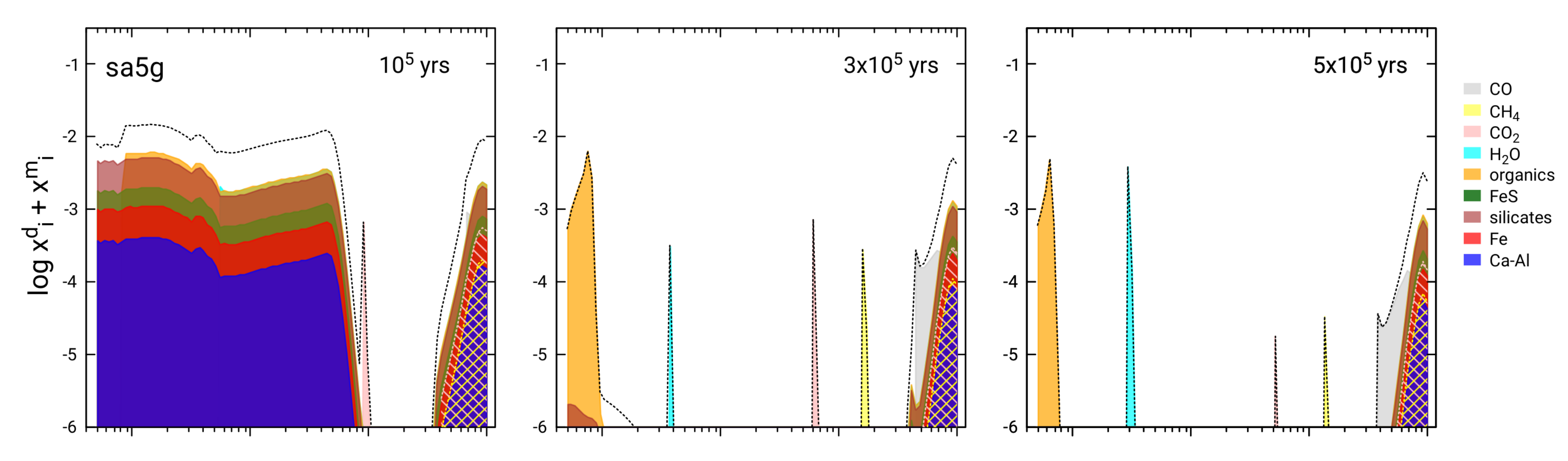}
\includegraphics[width=1.0\textwidth]{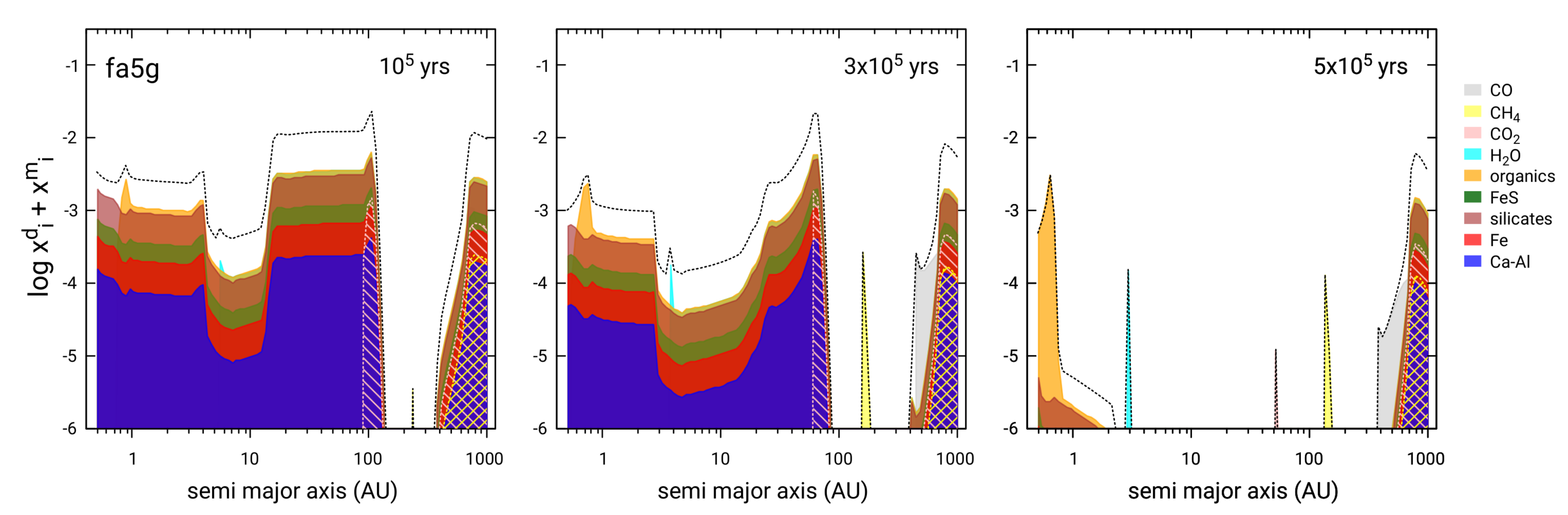}
\caption{Simulations with $\alpha_{\rm{t}}=10^{-5}$ of the evolution of solid fractions of all species for the compact particle (sa5g) and fractal aggregate (fa5g) growth at 0.1, 0.3 and 0.5 Myrs. The black dotted curves show the total of all species and represents the instantaneous metallicity $Z$. For this very low value of turbulent intensity, the disk evolution is more rapid, such that even in the fractal aggregate model most of the disk material has been lost after 0.5 Myr. By then, both models are mostly just characterized by bands of trapped or stranded material.  
\label{fig:a5g}}
\end{figure}

{\it Temperature variations with $\alpha_{\rm t}$: } %\textcolor{red}{Should this paragraph be located back with figure 1?}
Amongst the above models, previously we mentioned that the fiducial model ($\alpha_{\rm t}=10^{-3}$)  remains the hottest after 0.5 Myr (Fig. \ref{fig:diskevol}, panel c), and at earlier times there is variation in the cooling rates of the disk amongst the different models. This is not difficult to understand. The $\alpha_{\rm{t}}=10^{-2}$ case is cooler than the fiducial model despite having smaller particles and aggregates, which increases the opacity (Fig. \ref{fig:diskevol}, panel e), simply due to the order of magnitude decrease in gas density by 0.5 Myr compared to $\alpha_{\rm{t}}=10^{-3}$ (Fig. \ref{fig:diskevol}, panel a). The compact particle growth case with $\alpha_{\rm{t}}=10^{-4}$ is only marginally hotter than the corresponding high turbulence case because even though there is so much more nebula gas (and much more solids mass) for $\alpha_{\rm{t}}=10^{-4}$, the opacity has decreased significantly due to rapid particle growth. The fractal case for the low turbulence value $\alpha_{\rm{t}}=10^{-4}$  stays hotter longer than the compact growth case for this $\alpha_{\rm{t}}$, as seen in Fig. \ref{fig:a4g} (4th row) where the silicate EF persists up to 0.3 Myr, whereas it has already evolved inward of our computational inner boundary in the compact particle growth case by 0.1 Myr.  %\textcolor{red}{I rewrote some phrases in here to make the things being compared more clear.}

%\begin{figure}
%\includegraphics[width=1.0\textwidth]{sa5g_3P_sol.pdf}
%\includegraphics[width=1.0\textwidth]{fa5g_3P_sol.pdf}
% 
%\caption{Evolution of solid fractions of all species for the solid particle (sa5g) and fractal aggregate (fa5g) growth simulations with $\alpha_{\rm{t}}=10^{-5}$ at 0.1, 0.2 and 0.3 Myrs. The black dotted curves show the total of all species and represents the instantaneous metallicity $Z$.
%\label{fig:a5g}}
%\end{figure} 

{\it Extremely low $\alpha_{\rm t}$ case:} Finally, in Figure \ref{fig:a5g} we present models for $\alpha_{\rm{t}}=10^{-5}$. The compact (sa5g) and fractal (fa5g) cases  provide an even starker contrast to the  $\alpha_{\rm t} = 10^{-3}-10^{-2}$  models than did $\alpha_{\rm{t}}=10^{-4}$. For the compact particle growth case, such a low turbulent intensity leads to rapid loss of most solid material in a considerably shorter period of time than for $\alpha_{\rm{t}}=10^{-4}$. The end result is that after 0.3 Myr essentially only the banded EF-related structures remain, which only fade slightly after 0.5 Myr. The spike in H$_2$O ice at its EF actually increases slightly, due to condensation from a large reservoir of vapor phase water as the disk cools and the EF moves inwards. The fractal aggregate case fa5g  proceeds in a similar fashion to fa4g, with rapid growth outside the snowline to aggregate masses about an order of magnitude larger even than the $\alpha_{\rm{t}}=10^{-4}$ model ($\gtrsim 10^9$ g, see Fig. 10, \citetalias{Est21}), and correspondingly larger ${\rm{St}}$. The larger ${\rm{St}}$ quickly lowers the local $Z$ to values $\lesssim 0.001$. For this $\alpha_{\rm t}$, advection and diffusion by the nebula gas are of course very small. It is clear that even for this low $\alpha_{\rm{t}}$, the solid material in the fractal growth case persists longer than for the compact particle growth case, but unlike model fa4g, much of the mass in fa5g is gone by 0.5 Myr. However, we do find for this lowest  $\alpha_{\rm t}$ that there is a short period of time at $\sim 0.2$ Myr where the solids-to-gas ratio just outside the snowline in both growth models is greater than unity and, with both the mass-dominant compact  and aggregate particles having large ${\rm{St}}$, may briefly satisfy conditions for the streaming instability \citep[see][]{Umu20} and a burst of ice-rich planetesimal formation. This is discussed in detail in Section 4.1 of \citetalias{Est21}.

\section{Discussion} 
\label{sec:discuss}
 
%\subsection{Redistribution of Condensables}
%\label{sec:condense}

\subsection{Bulk Planetesimal Composition}
\label{sec:partcomp} 

Following the evolution of refractories and volatiles as they are transported throughout the nebula allows us to say something about what the bulk composition of a planetesimal might be if it formed at a particular location in the protoplanetary disk, and at a given time. A general idea can be gleaned from the various evolutionary plots in Sec. \ref{sec:evosolvap}, but here we generate a more direct visualization for specificity at 0.5 Myr, when silicates are condensed everywhere.   %In section \ref{sec:cavfwk} we discuss the limitations of the model. 
We caution again that our simulations do not include a chemical model, %\sout{so that these planetesimal compositions do not account for any complex chemistry that may occur in the disk.} \textcolor{red}{
but merely track our selected species as they evolve and change phase.

\begin{figure}
\gridline{\fig{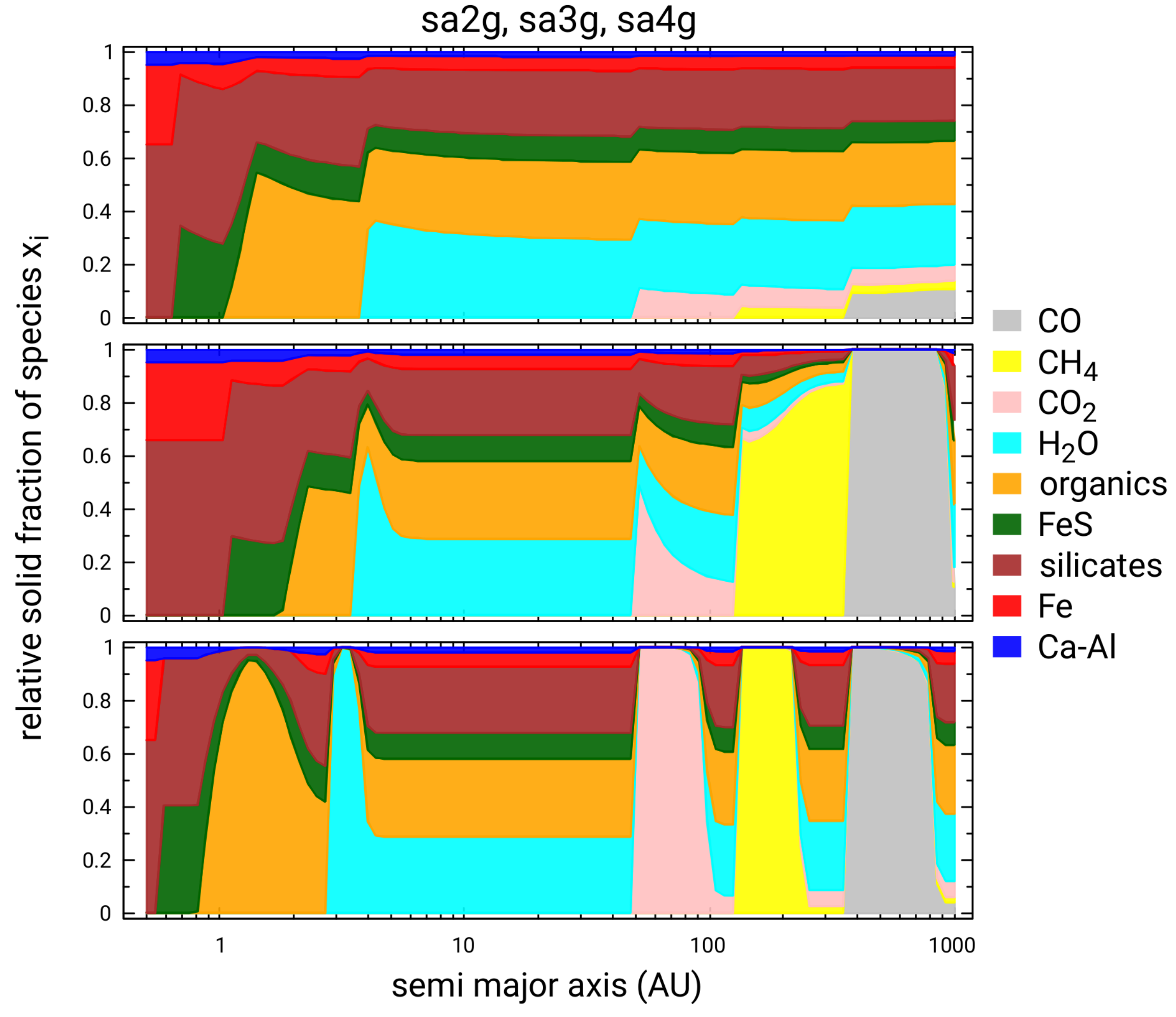}{0.48\textwidth}{}
          \fig{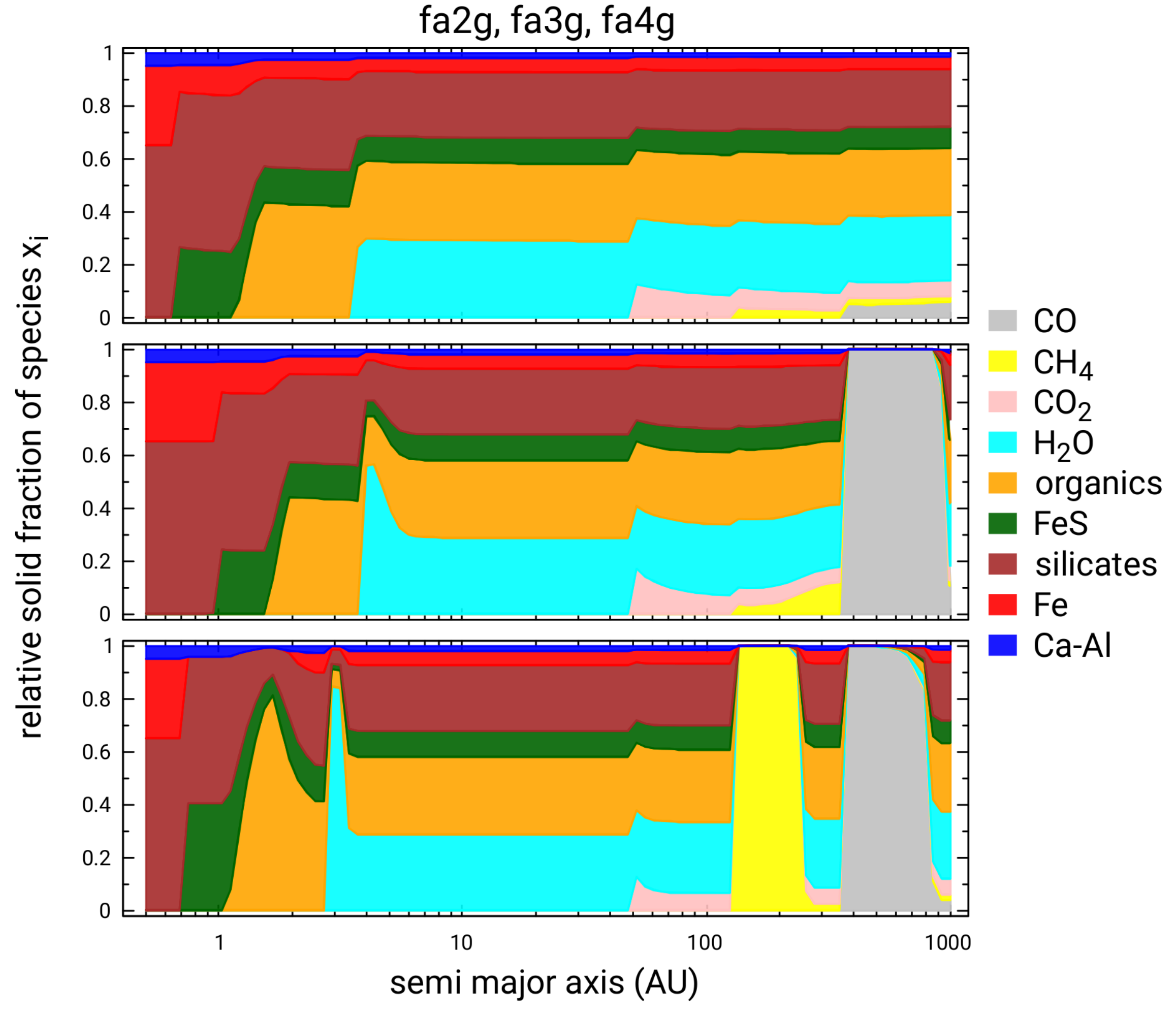}{0.48\textwidth}{}
          }
\caption{Fractional composition of compact particles (left panel) or fractal aggregates (right panel) after 0.5 Myr for simulations with $\alpha_{\rm{t}}=10^{-2}$ (top), $10^{-3}$ (middle) and $10^{-4}$ (bottom).  These panels represent the bulk composition a planetesimal might have if it formed at a given semi-major axis and does not take into account any pre- or post-accretional chemistry that may occur. The visualization is generated by cumulative summation at each radial location of all solid phase species.  %\textcolor{red}{figure caption should make it explicit that $\alpha = 10^{-2}$ is at the top, etc}
\label{fig:partcomp}}
\end{figure}

Figure \ref{fig:partcomp}  represents a snapshot of potential planetesimal composition  for different values of $\alpha_{\rm{t}}$. The left panel shows the compact growth models and the right panel the fractal aggregate growth models for turbulent intensities, from top to bottom, $\alpha_{\rm{t}} = 10^{-2}$, $10^{-3}$ and $10^{-4}$ (Table \ref{tab:models}). The locations of EFs, of which there are six still present in each simulation, are more easily seen in these figures as the radial location of the innermost boundary of a given species. The inner disk (region inside the snowline) is naturally the most refractory, but the variation in composition can be quite dramatic depending on $\alpha_{\rm{t}}$ and model type. In the outer nebula, temperatures are determined by stellar irradiation and not viscosity, so the locations of the EFs do not vary with $\alpha_{\rm t}$; the contents do, however. In the inner nebula, the fiducial case remains the hottest after this amount of time in the simulations (see Sec. \ref{sec:fiducial}), so even as far out as 1 AU, composition is dominated by silicates, iron and ``CAIs''. In the colder models at other $\alpha_{\rm t}$, or beyond 1 AU for $\alpha_{\rm{t}}=10^{-3}$, FeS contributes a  significant fraction to the bulk composition. As we explained in Sec. \ref{sec:varturb}, the models for $\alpha_{\rm{t}}=10^{-4}$ actually have no refractory iron over a wide region from $\sim$ 0.4AU - 2AU, as it has all been incorporated in troilite leaving a large reservoir of H$_2$S gas. It may be that this reservoir of H$_2$S gas goes on to affect mineralogy in other ways that are beyond the scope of this paper  \citep[see section \ref{sec:cavfwk}]{Lehneretal2013, Lehneretal2017}. Alternately, some of the iron could be sequestered in silicates. For example, our models initially begin with Mg-rich orthopyroxene (enstatite) and olivine (forsterite) silicates only, but once silicates are evaporated and recondense, the emergence of Fe-rich silicates (e.g., fayalite and ferrosilite) might be possible. As mentioned earlier, we have not included chemistry or mineralogy in our model yet  because it depends in a complicated way on the oxidation-reduction state of the local nebula, which can be influenced by the changing state or abundance of carbon as well as  by variation in the silicate and/or H$_2$O vapor abundance \citep[see also below and section \ref{sec:cavfwk}]{EbelGrossman2000,Tenneretal2015}. Further out still, ``moderately refractory" organics \citep{Pol94} %\textcolor{red}{Does Lodders discuss these? I thought not. I wish I could find my Grady reference.} 
which contain much of the C, can begin to dominate composition even creating a strong band of organics %\textcolor{red}{..of what..?} 
for the low turbulence case. This would suggest the possibility of very organic-rich bodies \citep[a ``tarline'', e.g., see][]{Lod04}, though as we have already mentioned, the enhancement of solid material outside the organics EF is likely overestimated in our current models because the organics probably break down into other constituents such as CO, NH$_3$ and/or hydrocarbons (e.g., C$_2$H$_2$) which would not recondense locally or in their ``moderately refractory" original form.

Just outside the H$_2$O snowline ($\sim 3-4$ AU in these models), there can be a large enhancement in water ice which can extend up to an AU or more outwards. If Jupiter's core formed at the snowline, this effect suggests that it might have benefited from a spike of enhanced water relative to silicates \citep[see also][]{KD19}. However, the composition of the Jovian ``core'' is not clear, and results from Juno imply that the atmosphere is depleted in water relative to other heavy element abundances, and could even be sub-solar \citep{Hel22}. %\textcolor{red}{Better see what the latest Juno results are and cite them. Also read and cite that Kalyaan and Desch paper on this topic.} 
Further outside the snowline out to about $\sim 50$ AU the composition is fairly uniform over the rest of the giant planet forming region (due to a lack of EFs there). Recall that this region is also expected to show a  systematic depletion of solids due to inward radial drift. The lack of EFs suggests that  the inwardly migrating particles may be able to drift relatively unaltered over much of this region, even as temperature increases. Jupiter is enhanced in C/H (as well as for noble gases and other volatiles) by a factor of $\sim 2-4$ \citep{Mah00,Won08,Bol17,LiC20}. One suggestion has been to attribute this enhancement to supervolatiles being trapped in crystalline ice as clathrate-hydrates \citep[e.g.,][]{Her08,Mou09}. An alternative is for supervolatiles to be trapped in amorphous ice in the colder regions of the disk \citep[e.g.,][also see \citealt{MD15}]{Bar07} that can be delivered to the giant planet region, but would have to be retained when amorphous ice transitions to crystalline \citep[e.g.,][]{Rua16}. The presence of Ar, which clathrates only at very low temperatures ($\sim 36$ K), %\textcolor{red}{.. clathrates ONLY at very low.... 36K?} 
as well as the presence of supervolatile ices on the giant planet moons, may favor the latter scenario \citep[e.g., see][for more discussion]{Est09}. %\sout{We will consider this in future modeling efforts.}

\begin{figure}
\gridline{\fig{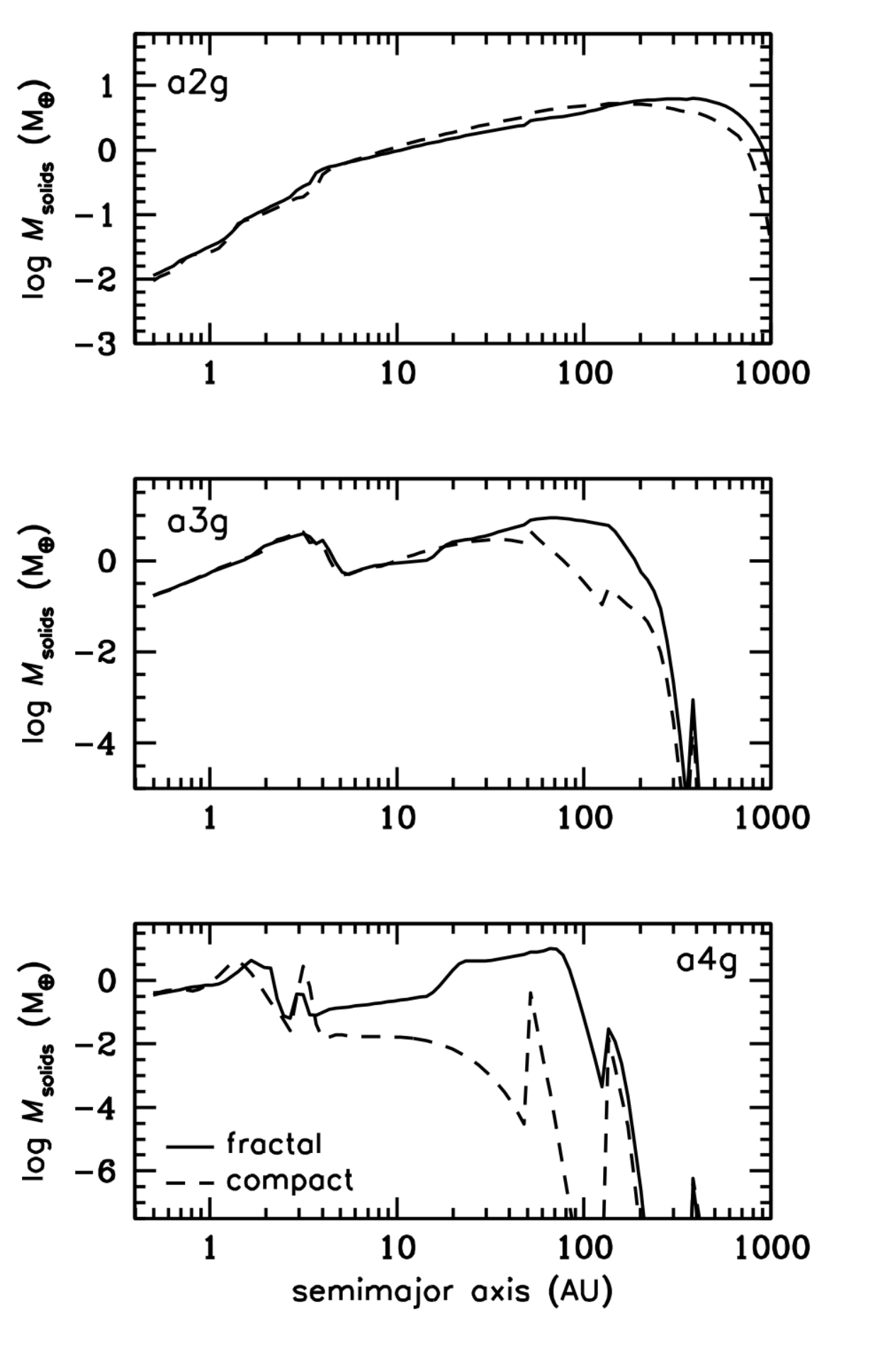}{0.48\textwidth}{}
          \fig{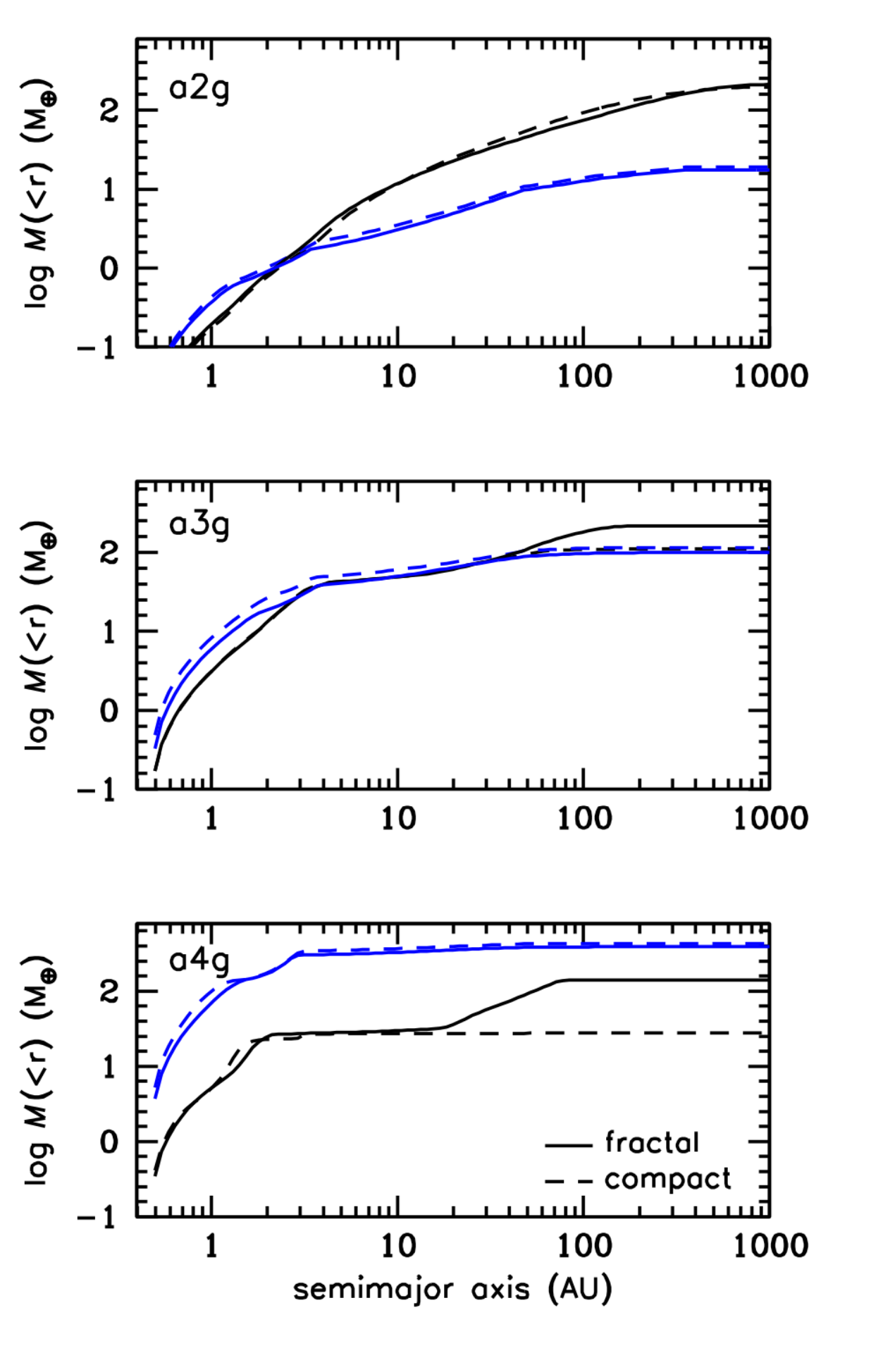}{0.48\textwidth}{}
          }
          \vspace{-0.5in}
\caption{Mass of solids in each radial bin  as a function of semimajor axis (left panels), and (right panels) cumulative mass of solids (black) and vapors (blue) in the disk for models of compact particles (dashed curves) and fractal aggregates (solid curves) after 0.5 Myr for simulations with $\alpha_{\rm{t}}=10^{-2}$ (top), $10^{-3}$ (middle) and $10^{-4}$ (bottom). Models a2g show that high $\alpha_{\rm{t}}$ leads to significant outward advection of solid material with the gas as the disk spreads. Lower $\alpha_{\rm{t}}$ models a3g and a4g are characterized by significant enhancement of material inside the snowline, but fractal models retain far more mass in the outer disk. In the case of a4g, most of the species mass is in the vapor phase and at small radii ($<3$ AU).
\label{fig:masscomp}}
\end{figure} 

Interestingly, in the very outermost portions of the disk, the supervolatile bands seen in models with $\alpha_{\rm{t}}=10^{-4}-10^{-3}$ suggests that it may be possible to have extremely supervolatile rich bodies that are essentially devoid of water forming in the coldest portions of the nebula, if planetesimals could form from the material ``trapped'' in these bands. Intriguing evidence along these lines may be found in the rare comets that are highly depleted in H$_2$O relative to CO, such as Comets C/2016 R2/PanSTARRS,  C/1908 R1 Morehouse, and C/1961 Humason \citep{Biver_etal_2018, Cochran_McKay2018, McKay_etal_2019}. In R2/PanSTARRS, CO/H$_2$O was enhanced by a factor of 1000 over normal comets. These condensation bands will continue to evolve inwards as the disk cools further, and advected nebula gas that still contains a vapor fraction of these supervolatiles could continue to replenish or even enhance these regions with time, but any planetesimals formed would remain where they formed. Recall that these regions have very little mass ($\Sigma_{\rm solids} \sim 10^{-2}$ g/cm$^2$, or even much lower), at least at this stage, though the model for $\alpha_{\rm{t}}=10^{-2}$ has advected a significant amount of material to these cold regions which is why the composition there is not extremely tilted in favor of the more volatile species.  

It is easier to visualize the possibilities by examining the amount of material available to form planetesimals at this epoch, which we show in Figure \ref{fig:masscomp}. In the left panel we show the total mass  %distribution 
of solids in each radial bin\footnote{We take the mass within a radial bin $R_j$ to be the solids surface density times the area of an annulus centered about $j$, $\pi \Sigma_{\rm{solids},j}(R^2_{j+1/2}-R^2_{j-1/2})$. In our code, we use logarithmic spacing for radial bins \citepalias{Est16}.} in terms of earth masses $M_\oplus$ as a function of semimajor axis (note the different scales covered for each $\alpha_{\rm{t}}$), while in the right panels we show the cumulative amount of solids (black curves) and vapor (blue curves) over the disk's radial extent. The top panels clearly demonstrate that the highest turbulent intensity model ($\alpha_{\rm{t}}=10^{-2}$) systematically transports material outwards %(roughly coinciding with the , \textcolor{red}{.. the what...???} 
while in the innermost regions where the gas velocity is inward, much of the depletion seen is due to material being lost to the star (left top panel), both in solid and vapor form. Indeed, the cumulative vapor fraction (right panels) is increasingly smaller than the cumulative solids fraction outside the snow line ($\sim 3-4$ AU), indicating that the enhancement (vaporization) at EFs has been muted due to both a lower radial drift mass flux, rapid outward advection and diffusion. Overall, for this $\alpha_{\rm{t}}$ there is as much solids mass inside $100$ AU ($\sim 100$ M$_\oplus$) as outside 100 AU, and it remains relatively uniformly mixed throughout the disk (Fig. \ref{fig:partcomp}, top panels).

On the other hand, lower turbulent intensities show distinct differences between the compact (dashed) and fractal models (solid). The compact growth models (dashed curves) lose more outer nebula  material via radial drift to inside the snowline with decreasing $\alpha_{\rm{t}}$ \citepalias{Est16}, while the fractal models retain much of the mass in the outer disk (beyond $\sim 20$ AU) for much longer timescales owing to the low ${\rm{St}}$, and thus slow radial drift times of the porous aggregates there \citepalias{Est21}. Radial drift leads to a large enhancement of solids for both the compact and fractal models in the inner disk inside the snowline with $\gtrsim 10$ M$_\oplus$ of material inside $\sim 2$ AU (whereas the $\alpha_{\rm{t}}=10^{-2}$ has $\lesssim 1$ M$_\oplus$), but the retention of material in the outer disk by the fractal models means that the total mass in solids out to $\sim 100$ AU for the fiducial ($\alpha_{\rm{t}}=10^{-3}$) model is $\sim 180$ M$_\oplus$, and $\sim 140$ M$_\oplus$ for $\alpha_{\rm{t}}=10^{-4}$ (compared to $\sim 100$ and $\sim 30$ M$_\oplus$ for the compact growth models, respectively). The retained solids mass in the outer nebula (that includes significant amounts of all species) and weaker radial drift in the fractal models also explains the lack of belts of CO$_2$ (and CH$_4$ for the fiducial model) seen in Fig. \ref{fig:partcomp}, which appear as notable bands in the compact growth model cases. Perhaps most striking is that most of the mass in the lowest $\alpha_{\rm{t}}$ model is in the vapor phase and close to the star, suggesting that this vast reservoir could eventually contribute to these bands as the disk cools and the vapor recondenses. Planetesimals that may form within these bands at an early time are resistant to inward migration as the EF itself evolves inwards, this may give rise to debris belts - ``fossils'' of a previous radial location of the EF - that may constantly produce pebbles and visible dust via collisions which may be reaccreted or drift inwards, as seen in the so-called ``Exo-Kuiper Belts" seen in exoplanetary systems \citep{Wyatt2020}.

%\textcolor{red}{Should calculate and state what the mass is in these bands, in $M_{\oplus}$. Also, comment in here somewhere that if actual planetesimals form within these bands at some early time (maybe we can get TH to give us a sanity check) they could resist inward migration as the ``stability zone" of the EF itself evolves inwards - becoming ``fossil" EFs, debris belts, etc, constantly producing ``pebbles" by collisions. Finally, something we really should comment on, is the vanishing of belts of first CO2-rich material and then CH4-rich material and then CO-rich material as $\alpha_{\rm t}$ is increased. It's very striking in figure 8, and they vanish first for the fractal cases - because more mass out there. Could this be a testable prediction of the models?  }

\subsection{Variation of Gas-phase C/O and C/H}
\label{sec:ctoo}

In the previous section, we made some inferences about the composition of a Jovian core depending on where it formed in the disk. The composition of the giant planet's envelope, and ultimately its atmosphere once runaway gas accretion begins \citep{Hub05,Lis09}, will also depend on the composition of the gas in the planet forming region in addition to subsequent planetesimal interactions with the envelope \citep[e.g.,][]{Pod20} and core-envelope interactions \citep[e.g.,][]{DC19,Liu19}. The composition of the nebula gas outside the snowline will be affected by the inward drift of compact particles and aggregates of higher volatility, as well as by outwardly advected and diffused small grains and vapor. As pointed out by \citet{Obe11}, because there would be large zones in the giant planet forming region where the nebula gas is enhanced in carbon with respect to oxygen but depleted with respect to hydrogen,  the C/O and C/H ratios in giant planet atmospheres can potentially elucidate details of their formation history, environment, and precursor planetesimals. 

\citet{Obe16} constructed simple models of the nebula in which they accounted for particle growth, settling and radial drift in an approximate way to show that %\textcolor{red}{the following should be rewritten as it is confusing, not sure what it means: .. Not sure why this is confusing. Right out of Oberg and Bergin 
one can get an excess of gas-phase C/H interior to the CO snow line (where ${\rm{C/O}}\sim 1$) due to the freezing out of carbon and oxygen-bearing species which drift across the CO EF and are sublimated. These authors concluded from their models that one might expect to find that gas giants that form in these types of environments would be enhanced over solar not only in C/O, but also in C/H - in contrast to their previous study \citep{Obe11}. Since in this work we model these processes explicitly, we may also look for these trends.

\begin{figure}
\centering
\includegraphics[width=0.4\textwidth]{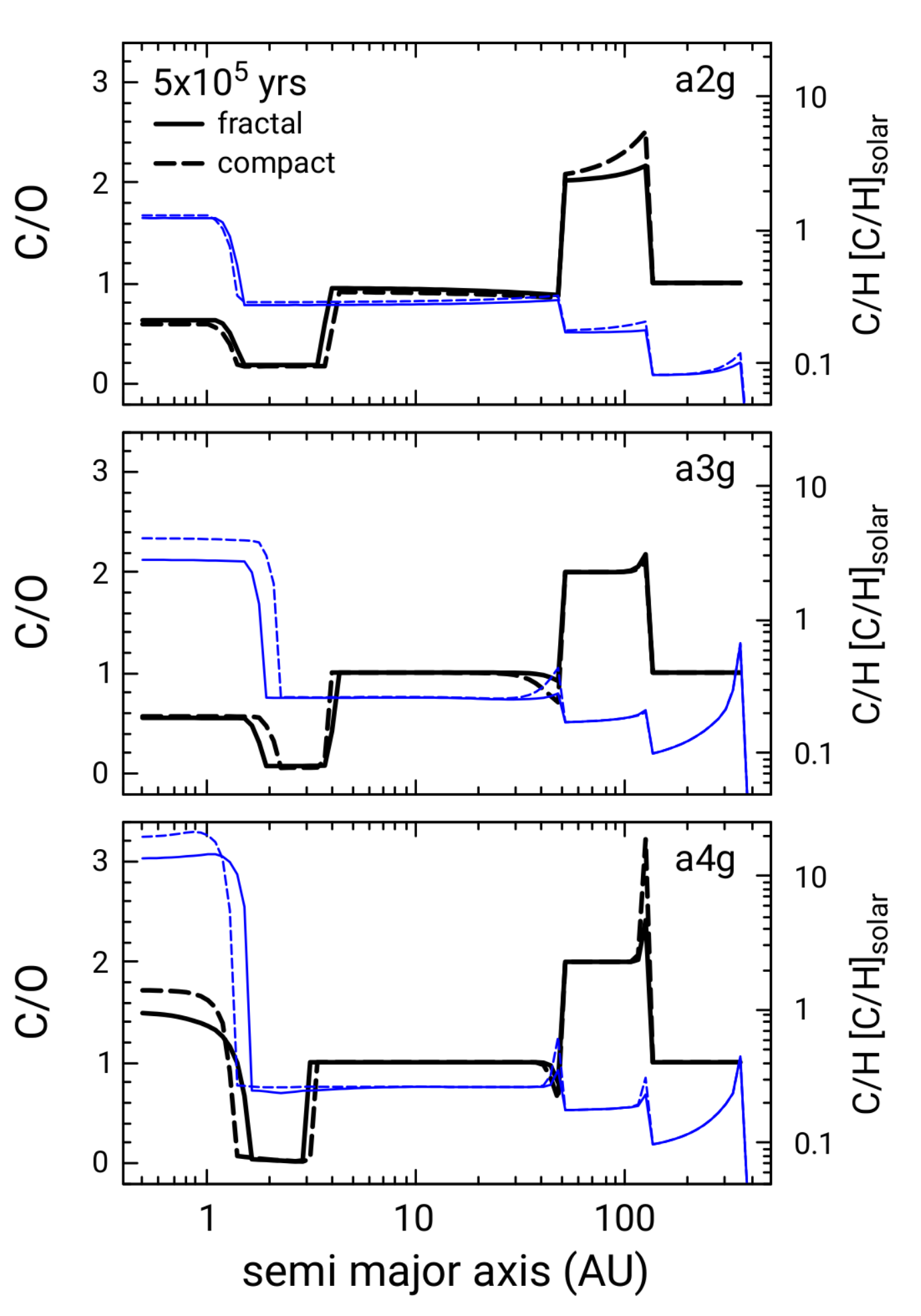}
\caption{Gas-phase C/O ratio (black curves) and C/H relative to solar (blue curves) as a function of semi-major axis for both fractal aggregate (solid curves) and compact growth (dashed curves) models after 0.5 Myr of evolution. Solar C/O = 0.5, and solar C/H = [C/H]$_{\rm{solar}} = 2.912\times 10^{-4}$ \citep{Lod03}.
%\textcolor{red}{here and elsewhere please clarify are BOTH the C/O and the C/H relative to their solar values? If not can you indicate what the solar valeus are in the plot or caption? Also what do you think about showing the Oberg et al results for comparison?} 
\label{fig:ctoo}}
\end{figure} 

In Figure \ref{fig:ctoo} we plot the gas-phase C/O ratio (black curves) which has a value of 0.5 for solar abundance  \citep{Lod03}, and C/H (blue curves) relative to solar ratio after 0.5 Myr for both fractal (solid curves) and compact particle (dashed curves) growth models with the different turbulent intensities shown. We note that these gas-phase profiles are generally weakly dependent on the growth model, or even the value of $\alpha_{\rm{t}}$. Most of the variation occurs at EFs, or in their vicinity.
%\textcolor{red}{OK?: Actually, these profiles are only weakly dependent on either $\alpha_{\rm t}$ or fractal/solid growth.} 
The various changes seen in the C/O ratio mark the CO EF ($\sim 350$ AU), methane EF ($\sim 150$ AU), CO$_2$ EF ($\sim 50$ AU), water ice EF ($\sim 3-4$ AU), and the organics EF ($\sim 1.5-2$ AU). The C/O ratio is roughly unity inside the CO snowline,  %\textcolor{red}{here it is again. Is C/O NOT normalized to solar but C/H is?this is confusing..}, This is done I guess to have the ratio of C/H closer to values about unity. Apparently the standard is just to do C/O (and other ratios) not relative to solar.
but increases by an additional factor of about 2 over solar as expected at the CH$_4$ EF. More highly localized increases above this value at the CH$_4$ EF are due to varying amounts of radial drift and rates of diffusion across the EF, that lead to enhancement of CH$_4$ vapor there. Once the CO$_2$ EF is reached, the C/O ratio returns close to unity except locally where excess CO$_2$ vapor  enhancement can cause ${\rm{C/O}}\lesssim 1$. Inside the H$_2$O snow line, the ratio drops considerably as water is sublimated, only to increase again substantially as refractory carbon is released into the gas-phase at the organics (tar line) EF.

The story for C/H proceeds somewhat similarly from the outer nebula inwards with subtle differences. We do see the trend of buildup of excess C/H inside the CO EF, and also at the methane and CO$_2$ EFs though less pronounced, largely in agreement with the models of \citet{Obe16}. We do not achieve the large $\gtrsim 3-10$ enhancement factors above solar C/H ($\sim 2.9\times 10^{-4}$, \citealt{Lod03}), but this is for a few reasons in addition to our models having a significant amount of C in refractory organics. First, these authors use a disk temperature profile that places the CO snowline at $\sim 50$ AU where there is considerably more solids material to contribute to the C/H excess. Second, the dynamical times there are much shorter so there has been more time in a relative sense for material to radially drift, sublimate and diffuse and recondense. %Third, our models include a significant fraction of C in refractory organics. % which likely would be broken down at the organics EF altering these profiles further.
%. Recall as mentioned in Sec. \ref{sec:varturb} (and below), refractory organics will more likely breakdown into simpler constituents like CO which would alter these profiles further, and over time. 
Lastly, the model of \citet{Obe16} considers only CO, CO$_2$ and water, and does not include CH$_4$ which allows C to evolve in a carrier that is not related to O. Thus it should not be surprising that different ratios are achieved that may partly be based on model assumptions, and not the process involved.
We expect that we would likely achieve large enhancements above unity under similar circumstances.
%\textcolor{red}{I also bet they do not have the large component of ``moderately refractory" organics that we do. Also, maybe somewhere in here, reiterate the worry about the role of reversible evaporation of ``moderately refractory" organics, it would seem to matter outside the tarline.}

Where we do see large excesses in C/H, of $\sim 1-20$x solar, is inside the tar line. As we have mentioned already, much of the refractory carbon may be irreversibly transformed into CO which can diffuse back across the organics EF, which would then lessen the inner nebula C/H excess somewhat, but this CO component likely would not diffuse much beyond $\sim 5$ AU over the simulation times here. The simple-minded view from these simulations is that a hot Jupiter forming inside the organics EF would  have its atmosphere enhanced in CO and H$_2$O \citep[e.g.,][]{for10}, and perhaps H$_2$S. On the other hand, a Jupiter forming just outside the water ice EF would have supersolar C/O, and perhaps remain subsolar in C/H depending on disk properties, but the C/H relative to solar should increase with radial distance in the disk out to the CO snowline. We note that the large C/H excesses inside the tarline in the vapor might have redox implications for meteoritic components - mineral stability and condensation temperature - that are beyond the scope of this paper.

\subsection{Caveats and Future Work}
\label{sec:cavfwk}

In our evolution code we employ a simplified approach to the treatment of EFs by allowing the local fractional abundance of a species to evaporate or condense linearly over a small temperature range about its associated condensation temperature \citepalias[see][]{Est16}. This phase change implementation avoids abrupt changes in opacity which may lead to unrealistic drops in temperature, and is meant to mimic the condensation process in which material first evaporates at the midplane, and then at higher altitudes in the disk with decreasing distance from the central star. This physical effect can buffer the midplane temperature near a constant value for some radial distance inwards, until the disk photosphere warms to the EF temperature. %; in the code, we merely impose this profile. NOT SURE WHAT THIS MEANS} 
We have found this approach works well for compact particles that are not too large, but for larger particles such as the especially fluffy porous aggregates that we find in our fractal growth models \citepalias{Est21}, it probably does not capture a more realistic scenario in which an aggregate's surface layers may insulate volatile material in their interiors. Under such circumstances, aggregates may be able to drift further past an EF releasing their associated volatile content more slowly. %This would likely have a smoothing effect over the region in which, for instance at the water snowline, not lead to as sharp a drop in aggregate mass with distance. Compositionally, 
Moreover, volatile material could perhaps even be retained as inward drifting aggregates sweep up layers of more refractory material. Thus an improved model would  treat the kinetics of evaporation and condensation at EFs \citep[e.g., see][]{RJ13,SO17}, which we will implement in future work. %\textcolor{red}{Better look at, and probably cite, Ros and Johansen and Schoonenberg and Ormel. Cuzzi et al and CC do not have especially sophisticated kinetics but these people do. }

For simplicity, we have also assumed that evaporation and condensation are reversible processes, but this is almost certainly not the case for all species. We have noted in this paper very large enhancements in both solids and vapor at the organics EF. However, we'd expect that these refractory organics would more realistically break down irreversibly into CO, CO$_2$, NH$_3$ and ``carbon chains'' like C$_2$H$_2$ and would {\it not} simply recondense on the cold side of the organics EF. Thus the large enhancement seen in the solid phase just outside the ``tar line", for instance in Fig. \ref{fig:a4g} or Figure \ref{fig:partcomp}, would likely not occur. We have already relaxed this assumption in \citep{Sen21}. This would seemingly be more consistent with the observation that the most primitive carbonaceous chondrites contain only $\sim 10$\% of the carbon found in comets \citep[see][]{Cuz03, Woodwardetal2021}. On the other hand, organics may be stickier \citep[][though see \citealt{Bis20}]{Kou02,Hom19} than what we assume in this work, which might facilitate faster growth to larger sizes before encountering the organics EF, and/or also allow for retention of some organics inwards of the EF if a proper model for evaporation of a larger aggregate were implemented. Naturally, releasing more CO into the disk would also affect the distribution of the gas-phase C/O ratio.

We have also implicitly assumed that sublimation and condensation of FeS is completely reversible, based on experimental results that suggest that the reaction between H$_2$S and free Fe is fast and progresses in such a way that it may likely consume all available Fe \citep{Arm_1960}. More recent experimental work shows that how fast FeS can form depends on the temperature, Fe grain size and H$_2$S and H$_2$ partial pressures \citep{Lau96,Lau97,Lau98}; that is, it is kinetically limited to some degree. Models by \citet{Pas05} calculated an upper limit on the timescale for formation of FeS for different Fe grain sizes assuming that the reaction temperature ranges from $\sim 680$ K where FeS decomposes to $\sim 500$ K below which formation of FeS is taken to be kinetically inhibited. They find S condensation onto small refractory Fe grains (such as our monomers) is quite short in the inner nebula at $680$ K, but the timescale increases by two orders of magnitude at $500$ K. In our models, if we did allow for the cutoff at $500$ K, we believe the effect would be to decrease the width of any band of depleted Fe (see e.g., Fig. \ref{fig:a4g} where the band width extends as far out as $\sim 2$ AU where $T\sim 300-400$ K), and further out beyond this band we would likely see coexisting refractory Fe, FeS and gas phase H$_2$S, with the latter possibly diffusing further outward or being involved in different chemical reactions. We will explore this in future models. %\textcolor{red}{But, how far outward radially (in decreasing temperature) do your enhanced FeS corrosion belts extend now? Do they extend past 500K?
 
As discussed in \citetalias{Est21}, we have used the optical constants for (crystalline) silicates from \citet[][see also \citealt{Cuz14}]{Pol94} in our calculations of the Rosseland and Planck mean opacities. However, more recent applications \citep[e.g.,][]{Bir18} tend to use (amorphous) astronomical silicates \citep{Dra03} which are more absorbing (especially at shorter wavelengths, by up to an order of magnitude). We should thus consider these as well, because particle opacity determines the conditions under which several recently discovered hydrodynamical instabilities leading to sustained turbulence may operate \citep[see][]{LU19}. In our simulations, we regularly find Rosseland mean opacities of $\sim 1-10$ cm$^2$ g$^{-1}$ which may be close to the limiting opacities that allow for these mechanisms. As the disk evolves, opacities can remain high in the inner disk and decrease in the outer disk for compact growth models, but can be sustained at higher values in the fractal aggregate models. We are actively pursuing better constraints on how particle growth can affect these turbulent instabilities. %\textcolor{red}{I'm not sure the difference between our refractive indices and Birnstiel's would make much difference, for your macroscopic particle sizes and thermal wavelengths. The differences don't even seem to be that great even in the mm range.}

For this work, we have assumed a constant background gas opacity of $10^{-4}$ cm$^2$ g$^{-1}$ \citepalias[see][]{Est16}, and have not calculated the Rosseland and Planck gas opacities. We have included these in a followup paper \citep[ which also includes a model for disk winds,][]{Sen21} as a table lookup derived from the work of \citet{Fre14} and find that it does not add any overhead to our code. For our simulations here, it is probably not as important because there is never a situation where all species have evaporated (ie, where temperatures exceed 2000 K). However, in models where the inner boundary is closer to the star \citep{Sen21}, where almost all species are in the vapor phase (or for very high $\alpha_{\rm{t}}$), the gas opacity dominates.  The gas opacity can even be  higher than the solids opacity and able to produce higher inner nebula temperatures, leading to steep gradients in $\nu$, and  possibly outward mass flux which might create a cavity. Studying the potential for strong outflow events will be the subject of future work. For full consistency with the epoch of interest, we will  also incorporate infall into our code. During the infall phase there is the possibility for significant growth and redistribution  as the disk forms \citep[e.g., see][]{HG05,Vis09,YangCiesla2012,DD18,HN18}. %\textcolor{red}{Add Hueso and Guillot 2005, and Visser et al 2009. }
%meridional flow? Layered accretion?

Finally, we note that the potential for large enhancements in vapor of the various species means that the molecular weight of the gas might be significantly altered, which would affect the local sound speed, %\textcolor{red}{ wild speculation - maybe even the mach number of shocks, with chondrule heating implications?}, 
and thus the gas viscosity \citep[see, e.g.][]{SO17,Cha21}. In our simulations we include the vapor content in the evolution of the gas surface density, but we have thus far not included the molecular weight effect. We have found that for the lowest turbulent intensity models the enhancement in vapor phase can be quite large. Thus, a significant decrease in sound speed at EFs might lead to a pressure bump that can produce even stronger enhancements in solids, and expand the conditions under which, for example, leap-frog planetesimal formation mechanisms may be satisfied. Our code can easily handle such a modification and we will include it in future work.

% Variable luminosity?
  
% Temperatures may be different for these species, e.g. refractory Fe? Maybe look at Lodders. May redo these for Denbanjan's paper so could say that here if we do so.
 
% Also T range for the cold ice models? Just took this from the paper, but that is for room temperature I thought, so not clear what it is in these conditions..

\section{Summary}
\label{sec:sum}

In this paper, and in our companion paper \citepalias{Est21} we have conducted simulations that compare compact particle and fractal aggregate growth models in globally evolving early stage (class 0/I) protoplanetary disks. In this paper, we have examined a small subset of the parameter space available, in particular the effects of varying the turbulent intensity $\alpha_{\rm{t}}$ within the models, in studying how the bulk fractions of solids and condensables of multiple species are redistributed in the nebula over time. We  have shown how  global evolution, including growth and drift of solids, and various EFs, likely had a significant influence on the radial distribution and composition of the most primitive bodies. %We summarize the general results below.

We find that compared to compact particle growth, fractal aggregate growth models for $\alpha_{\rm{t}} \lesssim 10^{-3}$ are characterized by a precipitous drop in the local metallicity $Z$ outside the water snowline and extending over much of the outer planet forming region ($\lesssim 20$AU) much earlier in the simulations (e.g., compare models after 0.1 Myr).
%precipitous \textcolor{red}{... rapid? ie it is rapid in time, not precipitous in radius?} drop in the local metallicity $Z$ outside the snow line and extending over much of the planet forming region ($\lesssim 20$AU). 
%Strong radial drift is encountered sooner in the fractal aggregate  simulations than the solid particle models. 
This is due to the more rapid growth of the porous aggregates to very large masses (and sizes) owing to their enhanced cross sections coupled with the higher fragmentation threshold for water ice (though see below for ``cold ice"). As a result these aggregates have stronger radial drift %earlier on
which more quickly transports their material to the inner disk inside the snow line.  Outside $\sim 20$ AU, while growth remains more rapid compared to compact growth models, the Stokes numbers of these porous aggregates remain much smaller than for the compact particles, keeping drift speeds low, which allows for the retention of significant amounts of mass in the outer disk even after 0.5 Myr (see Fig. \ref{fig:diskevol}, panel f). %for longer periods. 
compact growth models are thus  characterized by a more uniform, systematic loss of material from the outer disk to the inner disk due to radial drift. This is discussed in more detail in \citetalias{Est21}. %\sout{, where we also discuss constraints from mm-cm wavelength observations on particle porosity.} 
    
The magnitude of turbulent intensity $\alpha_{\rm{t}}$ unsurprisingly has a significant effect on the amount of mixing and redistribution of solids and vapors in the disk over time. We find that on average our fiducial  $\alpha_{\rm{t}}=10^{-3}$ for either fractal or compact growth tends to lead to the largest buildup in solids in the inner disk (inside the snowline) that is retained at least until 0.5 Myr. The   $\alpha_{\rm{t}} = 10^{-2}$ model never sees significant inner disk buildup, while lower turbulent  intensities lead to loss of inner disk solids into the star at an increasingly rapid rate. For $\alpha_{\rm{t}}=10^{-2}$, particle masses (and even more so sizes, and thus ultimately  Stokes numbers) stay small enough that particles or aggregates remain well coupled to the nebula gas. This explains the relatively featureless solids profiles, and neither they nor the corresponding vapor profiles see much enhancement ({\it i.e.}, exceed  the initial $Z$ much) anywhere in the disk. This lack of enhancements at EFs in the   $\alpha_{\rm t} = 10^{-2}$ case is also a result of  strong radial diffusion and advection (nebula gas velocities are inwards inside of $\sim 20$ AU), which rapidly smooth out any features that develop, regardless of whether particles are fractal or compact. Indeed, the distinction between the two types of models appears to vanish with increasing turbulent intensity.

The opposite is true for the lowest turbulent intensities ($\alpha_{\rm{t}} \lesssim 10^{-4}$). Despite significantly more disk material being processed across the snowline and feeding  the inner disk region due to faster growth and stronger radial drifts in the outer disk, material in the inner disk also drains away quickly into the star because particles and aggregates can still grow quite large (having larger Stokes numbers). % even though their fragmentation threshold is lower. 
Because radial diffusion and advection is weaker, we see that the enhancement peaks in both solids and vapor are more prominent, and persist throughout the simulation. With decreasing turbulent intensity, the vapor enhancements continue to grow in magnitude inside their corresponding EFs because diffusion back across the EF is slow compared to the radial drift of solids. Thus a characteristic of the low turbulence models is that much of the disk mass of solids ends up in the gas phase, especially in the inner disk region. In the outer disk, the vapor fraction is higher (at least earlier on) in the compact particle growth models due to  strong radial drift everywhere, further emphasizing the distinction between fractal aggregate and compact particle growth models with decreasing $\alpha_{\rm{t}}$. 

The retention of more material in the inner disk region (and a significant amount in the outer disk for the fractal growth models) by the fiducial ($\alpha_{\rm{t}} = 10^{-3}$) model compared to other values of the turbulent intensity can be explained as follows. In the inner disk inside the H$_2$O snowline, particle or aggregate Stokes numbers are small enough that radial drift is weak (see Fig. \ref{fig:diskevol}, orange curves panel f), and the fiducial $\alpha_{\rm{t}}$ means advection is less influential than for higher values, even if particles are more well coupled to the gas. The fiducial model is also the hottest model (note that fractal aggregate models are hotter in general) after 0.5 Myr indicating that the opacity remains quite high (since less material is lost due to slower drift), and in fact is comparable to the $\alpha_{\rm{t}}=10^{-2}$ models inside the snow line (Fig. \ref{fig:diskevol}, panel e). However, the gas density of the $\alpha_{\rm{t}} = 10^{-2}$ model %\textcolor{red}{OK? Antecendents becoming unclear} 
is an order of magnitude smaller (Fig. \ref{fig:diskevol}, panel a)  due to vigorous disk evolution, so the disk temperature is considerably lower by 0.5 Myr. After the same amount of time, the compact particle growth fiducial model has seen a drop in metallicity in the outer disk by an order of magnitude, while the fractal aggregate growth model still maintains $Z$ comparable to the initial value beyond $\sim 40$ AU (some smaller, low Stokes number aggregates have been advected outwards with the gas, increasing the mass there). The latter effect is even more pronounced when considering the ``cold H$_2$O ice" models (Fig. \ref{fig:a3Qg}) because the fragmentation threshold drops to a much lower value (similar to silicates) not far beyond the snowline, so both the compact particle and fractal aggregate growth models are characterized by significant retained mass in the outer regions overall. %\textcolor{red}{some rewording of this paragraph for more clarity, I hope.}

 An especially notable characteristic of the evolution of these different models - especially in the lowest $\alpha_{\rm{t}}$ simulations - is that processing of material across EFs eventually leads to bands of trapped associated volatile-rich material of radially variable compositions that can persist to much later times even if most of the mass of solids in the outer nebula has drifted radially inwards. %\textcolor{red}{ this was confusing to me; diffusion and advection are slow for low alpha, why should the effect be MOST noticeable there? please clarify: 
Even for low $\alpha_{\rm{t}}$, vapor can continue to diffuse or advect across the EF from within and condense, but for the lowest turbulent intensities, this process can be slow. %In the outer disk where one finds the supervolatile EFs, the outward advection of gas can continue to allow for further condensation of the associated supervolatile with time. 
Steep gradients in vapor can build up with time inside the EF though which can help to facilitate diffusion (e.g., see Fig. \ref{fig:a4g}, panels 2 and 4) possibly helping to make the bands long-lived. As the outer disk cools (in our models, when the luminosity decreases significantly), these bands also would evolve further inwards also allowing for more vapor to condense.  Belts of planetesimals that may be formed there or at their earlier locations would remain in place (see below). The amount of mass in these bands at the epoch of our simulations is relatively small (up to $\sim 1$ M$_\oplus$ to $10^{-2}$ M$_\oplus$, or even smaller; %\textcolor{red}{give rough numbers, ??$M_{\oplus}$?} 
see also Fig. \ref{fig:diskevol}, panel b, and Fig. \ref{fig:masscomp}), %\textcolor{red}{I am not sure I saw a quote of $\Sigma$ anywhere for the individual bands} 
but this could change with longer term evolution. Recall that most condensible material mass is in the vapor phase closer to the sun for the lowest $\alpha_{\rm{t}}$ models (see Fig. \ref{fig:masscomp}). The EF-related banded structure, or enhancements outside the various ``snow lines'' in general, as well as significant radial dips in the metallicity seen in our models, may provide an  explanation for ALMA observations that show well defined particle belts at large radii  \citep{CG19,SC20}, or for the formation of ``ExoKuiper belts" seen in revealed systems \citep{Wyatt2020}.

Finally, %\sout{and related to the above,} 
the redistribution of solids and condensables with time provides some insight into the possible bulk composition as a function of semi-major axis of
%naturally leads to implications for the radially dependent distribution of 
planetesimals that may form through leapfrog processes like the streaming instability or turbulent concentration \citep[][see \citetalias{Est21}]{Joh14, Umu20, HC20}, as well as the enhancement in the atmospheres of giant planets. We find that unless the turbulent intensity is very high, planetesimals that form outside EFs will be more rich in the associated volatile. The implications can include that a Jovian core that forms at the snowline would be rich in water ice \citep{KD19}, or in refractory organics if it formed  further inwards,  near the tarline \citep{Lod04}. We are, however, skeptical about the latter possibility (see  above in this section). On the other hand, the water content in Jupiter continues to be uncertain, and remains the subject of investigation \citep{Hel22}. A more novel and intriguing possibility may be the formation of supervolatile-rich and water-ice-poor bodies outside of various EFs (Fig. \ref{fig:partcomp}). As one example, the distinct supervolatile belts might help explain the different compositions of different dynamical KBO populations \citep[e.g.,][]{Bro12,WB17}, including Arrokoth where water ice was not detected, but the presence of methanol on its surface may be attributed to hydrogenation of CO-ice \citep{Pen20,Gru20}.  As another example, these belts may also explain the CO-rich composition of comets like C/2016 R2/PanSTARRS \citep{Biver_etal_2018, Cochran_McKay2018,Wierzchos_Womack_2018, McKay_etal_2019}.
%A more novel and intriguing possibility may be the formation of supervolatile-rich bodies - comets and TNOs - outside of their respective EFs. The combinations of turbulent intensity and local particle mass density and St characterizing these models do not lead to planetesimal formation by SI in these regions \citep{Umu20}, but TC may remain a possibility \citep{HC20}. This might help explain the different compositions of different dynamical KBO populations  \citep[e.g.,][]{Bro12,WB17}, the CO-rich composition of comets like C/2016 R2/PanSTARRS \citep{Biver_etal_2018, Cochran_McKay2018,Wierzchos_Womack_2018, McKay_etal_2019}, or some KBOs such as Arrokoth where water ice was not detected, but the presence of methanol on its surface may be attributed to, for example, hydrogenation of CO-ice \citep{Pen20,Gru20}. 
Evolutionary effects in the nebula {\it gas} could also have important implications. Giant planets forming inside the snowline (hot Jupiters) may have had their atmosphere enhanced in both water and CO \citep{for10}, especially if we consider that the evaporation of refractory organics would place a lot of the C in CO. Some of our models also generate excess H$_2$S well inside the snowline. If the core formed outside the snow line, then its atmosphere may be supersolar in C/O.

%Hottest model is 1e-3 sweet spot? (generally fractal models hotter for lower alphas?)

%H2S FeS thing

\acknowledgments

We thank the NASA Ames Research Center's Origins Group, and specifically useful conversations with Uma Gorti,  Maxime Ruaud, Orkan Umurhan, Diane Wooden, Mike Kelley, and Kevin Zahnle. We thank an anonymous referee for very useful comments that helped to improve the clarity of the paper. P.R.E. and J.N.C. acknowledge support from the NASA Emerging Worlds Program grant NNX17AL60A and a NASA ISFM grant.

\bibliography{my.bib}



\end{document}